\documentclass[conference]{IEEEtran}

\usepackage{amsfonts}
\usepackage{amsmath} 
\usepackage{amssymb}  

\usepackage{float}

\usepackage{graphicx}
\usepackage{epstopdf}

\usepackage{amsthm}

\theoremstyle{definition}
\newtheorem{proposition}{Proposition}[section]

\usepackage{tikz}
\usetikzlibrary{calc,patterns,decorations.pathmorphing,decorations.markings}
\usetikzlibrary{shapes,arrows}
\usepackage{verbatim}

\tikzstyle{block} = [draw, fill=blue!20, rectangle, 
    minimum height=3em, minimum width=6em]
\tikzstyle{sum} = [draw, fill=blue!20, circle, node distance=1cm]
\tikzstyle{input} = [coordinate]
\tikzstyle{output} = [coordinate]
\tikzstyle{pinstyle} = [pin edge={to-,thin,black}]

\usetikzlibrary{positioning}
\usetikzlibrary{math}

\usepackage{tikz}
\usetikzlibrary{shapes,arrows,calc,positioning}
\tikzstyle{bigblock} = [draw, fill=blue!20, rectangle, 
    minimum height=6em, minimum width=8em]
\tikzstyle{medblock} = [draw, fill=blue!20, rectangle, 
    minimum height=4em, minimum width=4em]    
\tikzstyle{mux} = [draw, fill=black!20, rectangle, 
    minimum height=5em, minimum width=0.1em]    
\tikzstyle{smallblock} = [draw, fill=blue!20, rectangle, 
    minimum height=3em, minimum width=4em]
\tikzstyle{sum} = [draw, fill=blue!20, circle, node distance=1cm]
\tikzstyle{signal} = [coordinate]
\tikzstyle{pinstyle} = [pin edge={to-,thin,black}]
\tikzstyle{block} = [draw, fill=blue!20, rectangle, 
    minimum height=3em, minimum width=6em]
\tikzstyle{blockS} = [draw, fill=blue!20, rectangle, 
    minimum height=3em, minimum width=4em]    
\tikzstyle{input} = [coordinate]
\tikzstyle{output} = [coordinate]
\usetikzlibrary{matrix}

\newcommand{\bc}{\begin{center}}
\newcommand{\ec}{\end{center}}
\newcommand{\benum}{\begin{enumerate}}
\newcommand{\eenum}{\end{enumerate}}
\newcommand{\nn}{\nonumber}
\newcommand{\matl}{\left[ \begin{array}}
\newcommand{\matr}{\end{array} \right]}
\newcommand{\matls}{\left[ \begin{smallmatrix}}
\newcommand{\matrs}{\end{smallmatrix} \right]}
\newcommand{\isdef}{\stackrel{\triangle}{=}}

\newcommand{\vect}[1]{\overset{\rightharpoonup}{#1}}

\newcommand{\rmE}{{\rm E}}
\newcommand{\rmF}{{\rm F}}

\newcommand{\rmP}{{\rm P}}
\newcommand{\rmQ}{{\rm Q}}

\newcommand{\rmT}{{\rm T}}

\newcommand{\rmW}{{\rm W}}

\newcommand{\rmc}{{\rm c}}
\newcommand{\rmd}{{\rm d}}

\newcommand{\rmi}{{\rm i}}

\newcommand{\rmp}{{\rm p}}

\newcommand{\BBR}{{\mathbb R}}

\usepackage{subfig}   
\title{An A Quadcopter Autopilot Based on an Adaptive Digital PID Controller}

\title{Adaptive Digital PID Control of a Quadcopter}

\title{Retrospective-Cost-Based Adaptive Digital PID Control of a Quadcopter}

\title{A Retrospective-Cost-Based Adaptive Digital PID   Quadcopter Autopilot}

\title{Adaptive Digital PID Control of a Quadcopter with Unknown Dynamics}

\author{
    Ankit Goel, 
    Abdulazeez Mohammed Salim, 
    Ahmad Ansari, 
    Sai Ravela, 
    Dennis Bernstein
\thanks{Ankit Goel, Ahmad Ansari, and Dennis Bernstein are with the Department of Aerospace Engineering, University of Michigan, Ann Arbor, MI 48109.
{\tt\small ankgoel@umich.edu}
{\tt\small ansahmad@umich.edu}
{\tt\small dsbaero@umich.edu}} 
\thanks{Abdulazeez Mohammed Salim is with the Department of Aeronautics and Astronautics, MIT, Cambridge, MA 02139.
{\tt\small azez@mit.edu}} 
\thanks{Sai Ravela is with the Department of Earth, Atmospheric, and Planetary Sciences, MIT, Cambridge, MA 02139.
{\tt\small ravela@mit.edu}} 
}

\date{August 2019}

\begin{document}

\maketitle
\begin{abstract}
This paper develops an adaptive autopilot for  quadcopters with unknown dynamics.
To do this, the PX4 autopilot architecture is modified so that the feedback and feedforward controllers are replaced by  adaptive control laws based on retrospective cost adaptive control (RCAC).
The present paper provides a numerical investigation of the performance of the adaptive autopilot on a quadcopter with unknown dynamics.
In order to reflect the absence of prior modeling information, all of the adaptive digital controllers are initialized at zero gains.
%
%
In addition, moment-of-inertia of the quadcopter is varied to test the robustness of the adaptive autopilot. 
%
%
In all test cases, the vehicle is commanded to follow a given trajectory, and the resulting performance is examined.
\end{abstract}

\section{Introduction}

Multicopters are ubiquitous and are increasingly used for diverse applications ranging from sports broadcasting to wind-turbine inspection \cite{chang2016development,anweiler2017multicopter,andaluz2015nonlinear, schafer2016multicopter,stokkeland2015autonomous}.
In the simplest configuration, differential torques applied to the motors of a quadcopter provide thrust for translational motion as well as moments for attitude control.
For commercial applications, the autopilot of a quadcopter can be finely tuned and tailored to the geometry and mass properties of the vehicle.
In fact, the open-source autopilot PX4 has been used extensively for many vehicle configurations \cite{meier2015px4}.


In some applications, however, the vehicle properties are frequently modified due to changes in the airframe, payload, sensors, and actuators.
This occurs especially in experimental situations and field operations.
In these cases, there is no guarantee that stock autopilot gains will perform in an acceptable manner.
Along the same lines, unanticipated and unknown changes that occur during flight due to failure or damage may significantly degrade the performance of the autopilot.

With this motivation in mind, the present paper develops an adaptive autopilot for quadcopters with unknown dynamics.
To do this, the PX4 autopilot architecture is modified so that the feedback and feedforward controllers are replaced by  adaptive control laws based on retrospective cost adaptive control (RCAC) \cite{rahmanCSM2017}.
In particular, each PID controller in PX4 is replaced by an adaptive digital PID controller as described in \cite{rezaPID}.
The adaptive digital PID controller is based on recursive least squares (RLS), and thus involves the update of a matrix of size upto $4\times4$ at each time step, which is amenable to real-time implementation on a typical embedded processor used to support the PX4 autopilot.

Fuzzy neural network based sliding mode control was used in \cite{kayacan2017learning} to control a UAV in the presence of wind which learned the inverse dynamics of the plant model.
However, the autopilot needed P controllers to be suitably initialized to provide sufficient time for learning.
In contrast, the adaptive autopilot controllers are initialized at zero in this paper. 
Retrospective-cost based PID controllers were used in the attitude controller in \cite{ansari2018retrospective}, and were applied with fixed hyperparameters tuning to a quadcopter, a fixed-wing aircraft, and a VTOL aircraft. 
The present paper extends the work in \cite{ansari2018retrospective} by replacing all of the controllers in PX4 autopilot with adaptive controllers.

The contribution of the present paper is the development and numerical demonstration of an adaptive digital autopilot for poorly modeled quadcopters.
In particular, the present paper provides a numerical investigation of the performance of the adaptive digital PID autopilot on a quadcopter with unknown dynamics. 
%
In order to reflect the absence of prior modeling information, all of the adaptive digital controllers gains are initialized at zero. 
Next, the effect of tuning hyperparameters of the adaptive digital controllers on the closed-loop performance is investigated. 
Finally, the moment-of-inertia of the quadcopter is scaled by a factor of five and the adaptive autopilot with fixed tuning hyperparameters is applied to follow a given trajectory. 
%
%
%
%
In addition, the evolution of the adaptive controller gains is examined in order to compare the converged controller gains to the stock gains.


The paper is organized as follows.
In section \ref{sec:QuadDyn}, the quadcopter dynamics is summarized. 
In section \ref{sec:PID_Algo}, the retrospective cost based adaptive PID control algorithm is presented. 
In section \ref{sec:AdaptiveAutopilot}, the control architectures of stock PX4 autopilot and the adaptive PX4 autopilot are presented.
In section \ref{sec:Numerical}, simulation results are presented to compare the performance of the adaptive PX4 autopilot with the stock PX4 autopilot.
Finally, section \ref{sec:conclusions} concludes the paper with the summary of the paper and future directions.

\section{Quadcopter Dynamics}
\label{sec:QuadDyn}
The Earth frame and quadcopter body-fixed frame are denoted by the row vectrices $\rm{F}_{\rm{E}} = \begin{bmatrix} \hat{\imath}_{\rm{E}} & \hat{\jmath}_{\rm{E}} & \hat{k}_{\rm{E}} \end{bmatrix}$ and $\rm{F}_{\rm{Q}} = \begin{bmatrix} \hat{\imath}_{\rm{Q}} & \hat{\jmath}_{\rm{Q}} & \hat{k}_{\rm{Q}} \end{bmatrix}$, respectively. 
We assume that $\rm{F}_{\rm{E}}$ is an inertial frame and the Earth is flat. 
The origin $\rm{O_E}$ of $\rm{F}_{\rm{E}}$ is any convenient point fixed on the Earth.
The axes $\hat\imath_{\rm{E}}$ and $\hat\jmath_{\rm{E}}$ are horizontal, while the axis $\hat{k}_{\rm{E}}$ points downward.
%
%
$\rm{F}_{\rm{Q}}$ is defined with $\hat\imath_{\rm{Q}}$ and $\hat\jmath_{\rm{Q}}$ in the plane of the rotors, and $\hat{k}_{\rm{Q}}$ points downward, that is, $\hat{k}_{\rm{Q}} = \hat\imath_{\rm{Q}} \times \hat\jmath_{\rm{Q}}$. 
Assuming that $\hat\imath_{\rm{E}}$ points North and $\hat\jmath_{\rm{E}}$ points  East, it follows that the Earth frame is a local NED frame.
The quadcopter frame $\rm F_{\rm Q}$ is obtained by applying a 3-2-1 rotation sequence to the Earth frame $\rm F_{\rm E}$, where the 3-2-1 Euler angles $\Psi, \Theta, \Phi$ denote yaw, pitch, and roll angles, respectively.


The translational equations of motion of the quadcopter are given by
\begin{align}
    m \overset{\rmE \bullet \bullet} { \vect r}_{\rmc/\rm{O_E}} 
        &=
            m \vect g + \vect f,
    \label{eq:N2L_pos}
\end{align}
where 
$m$ is the mass of the quadcopter,
$\rmc$ is the center-of-mass of the quadcopter, 
$\vect r_{\rmc/\rm{O_E}} $ is the physical vector representing the position of the center-of-mass of the quadcopter relative to $\rm O_E$, 
$\vect f = f_z \hat k_{\rm Q} $, and 
$\vect g = g \hat k_\rmE$.
Let
\begin{align}
    \vect r_{\rmc/\rm{O_E}} 
        &=
            X \hat \imath_\rmE +
            Y \hat \jmath_\rmE +
            Z \hat k_\rmE,
    \label{eq:Pos_in_E}
        \\
    \overset{\rmE \bullet} { \vect r}_{\rmc/\rm{O_E}} 
        &=
            U \hat \imath_{\rm Q} +
            V \hat \jmath_{\rm Q} +
            W \hat k_{\rm Q}.
    \label{eq:Vel_in_E}
\end{align}

It thus follows from \eqref{eq:N2L_pos}-\eqref{eq:Vel_in_E} that
\begin{align}
    \dot{X} 
        &= 
            (\cos\Theta)(\cos\Psi) U + [ (\sin\Phi)(\sin\Theta)\cos\Psi \nn \\ 
            & \quad\,\, -(\cos\Phi)\sin\Psi ] V + [ (\cos\Phi)(\sin\Theta)\cos\Psi \nn \\
            & \qquad\qquad\qquad\qquad\qquad\quad +(\sin\Phi)\sin\Psi ] W, \\
    \dot{Y} 
        &= 
            (\cos\Theta)(\sin\Psi) U + [ (\sin\Phi)(\sin\Theta)\sin\Psi \nn \\ 
            & \quad\,\, +(\cos\Phi)\cos\Psi ] V + [ (\cos\Phi)(\sin\Theta)\sin\Psi \nn \\
            & \qquad\qquad\qquad\qquad\qquad\quad -(\sin\Phi)\cos\Psi ] W, \\
    \dot{Z} 
        &= 
            -(\sin\Theta) U + (\sin\Phi)(\cos\Theta) V + (\cos\Phi)(\cos\Theta) W, \\
    \dot{U} 
        &=
            VR -WQ -(\sin\Theta)g 
            ,\\
    \dot{V} 
        &=
            -UR +WP +(\sin\Phi)(\cos\Theta)g
            ,\\
    \dot{W} 
        &= UQ -VP +(\cos\Phi)(\cos\Theta)g + \frac{f_z}{m}.
\end{align}

Neglecting the gyroscopic moments due to the rotors' inertia, drag forces, and moments, the rotational equations of motion of the quadcopter in coordinate-free form are given by
\begin{align}
    \vec J_{{\rm Q}/\rmc} 
    \overset{\rm\rmE \bullet } { \vect \omega}_{\rmQ/\rmE} +
    \vect \omega_{\rmQ/\rmE} \times \vec J_{{\rm Q}/\rmc} \vect \omega_{\rmQ/\rmE}
        &=
            \vect M_{{\rm Q}/\rmc},
    \label{eq:EulersEqn}
\end{align}
where 
$\vec J_{{\rm Q}/\rmc} 
    =
        J_{xx} \hat \imath_{\rm Q} \hat \imath _{\rm Q}' +
        J_{yy} \hat \jmath_{\rm Q} \hat \jmath _{\rm Q}' +
        J_{zz} \hat k_{\rm Q} \hat k_{\rm Q}'$
is the inertia tensor of the quadcopter,  
$\vect M_{{\rm Q}/\rmc}$ is the moment applied to the quadcopter relative to $\rmc$, and 
$\vect \omega_{\rmQ/\rmE} 
    = 
        P \hat \imath _{\rm Q} + 
        Q \hat \jmath _{\rm Q} + 
        R \hat k _{\rm Q}$
        is the angular velocity of frame $\rmF_\rmQ$ relative to the inertial Earth frame $\rmF_\rmE.$ 
It follows from \eqref{eq:EulersEqn} that
\begin{align}
    \dot{\Phi} 
        &=
            P + (\sin\Phi)(\tan\Theta) Q + (\cos\Phi)(\tan\Theta)R, \\
    \dot{\Theta} 
        &= 
            (\cos\Phi) Q - (\sin\Phi)R, \\
    \dot{\Psi} 
        &=
            (\sin\Phi)(\sec\Theta) Q + (\cos\Phi)(\sec\Theta) R, \\
    \dot{P} 
        &=
            \frac{1}{J_{xx}} \left[(J_{yy}-J_{zz})QR  +  M_x    \right]  , \\
    \dot{Q} 
        &=
            \frac{1}{J_{yy}} \left[ (J_{zz}-J_{xx})PR + M_y \right]  , \\
    \dot{R} 
        &=
            \frac{1}{J_{zz}} \left[(J_{xx}-J_{yy})PQ + M_z    \right] .
\end{align}

\section{Adaptive Digital PID Control Algorithm}
\label{sec:PID_Algo}

The quadcopter is controlled by a digital controller operating in a sampled-data feedback loop.
%
%
%
In particular, consider the  PID controller  
\begin{align}
    u_k
        =
            K_{\rmp,k} z_{k-1} +
            K_{\rmi,k} \gamma_{k-1} +
            K_{\rmd,k} (z_{k-1} - z_{k-2}),
    \label{eq:uk_PID}
\end{align}
where $K_{\rmp,k}, K_{\rmi,k}, K_{\rmd,k}$ are time-varying gains to be adapted, $z_k$ is an error variable, and, for all $k\ge0$,
\begin{align}
    \gamma_k 
        \isdef
            \sum_{i=0}^{k} z_{i}.
\end{align}
Note that the integrator state can be computed recursively using $\gamma_k = \gamma_{k-1} + z_{k}$.
Finally, note that the control \eqref{eq:uk_PID} can be written as
\begin{align}
    u_k 
        =
            \phi_k \theta_k,
    \label{eq:uk_reg}
\end{align}
where, for all $k\ge0$,
\begin{align}
    \mspace{-8mu}\phi_k
        \isdef
            [
                z_{k-1} \ \
                \gamma_{k-1} \ \
                z_{k-1} - z_{k-2}
            ], \quad
    \theta_k
        \isdef
            \matl{c}
                K_{\rmp,k} \\
                K_{\rmi,k} \\
                K_{\rmd,k}
            \matr.
    \label{eq:phi_theta_def}
\end{align}

To determine the controller gains $\theta_k$, let $\theta \in \BBR^3$, and consider the \textit{retrospective performance variable} defined by
\begin{align}
    \hat{z}_{k}(\theta)
        \isdef
            z_k + 
            \sigma (\phi_{k-1} \theta - u_{k-1}),
    \label{eq:zhat_def}
\end{align}
where $\sigma$ is either $1$ or $-1$ depending on whether the sign of the leading numerator coefficient of the transfer function from $u_k$ to $z_k$ is positive or negative, respectively.
Furthermore, define the \textit{retrospective cost function} $J_k \colon \BBR^3 \to [0,\infty)$ by
\begin{align}
    J_k(\theta) 
        \isdef
            \sum_{i=0}^k
                \hat{z}_{k}(\theta)^2 +
                (\theta-\theta_0)^\rmT 
                P_0^{-1}
                (\theta-\theta_0),
    \label{eq:RetCost_def}
\end{align}
where $\theta_0\in\BBR^3$ is the initial vector of PID gains and $P_0\in\BBR^{3\times 3}$ is positive definite.
For all examples in this paper, we set $\theta_0 = 0$; however, $\theta_0$ can be initialized to nonzero gains in practice if desired.

\begin{proposition}
    Consider \eqref{eq:uk_reg}--\eqref{eq:RetCost_def}, 
    where $\theta_0 \in \BBR^3$ and $P_0 \in \BBR^{3 \times 3}$ is positive definite. 
    Furthermore, for all $k\ge0$, denote the minimizer of $J_k$ given by \eqref{eq:RetCost_def} by
    \begin{align}
        \theta_{k+1}
            \isdef
                \underset{ \theta \in \BBR^n  }{\operatorname{argmin}} \
                J_k({\theta}).
        \label{eq:theta_minimizer_def}
    \end{align}
    Then, for all $k\ge0$, $\theta_{k+1}$ is given by 
    \begin{align}
        \theta_{k+1} 
            &=
                \theta_k  + 
                 P_{k+1}\phi_{k-1}^\rmT
                 [ z_k + \sigma(\phi_{k-1} \theta_k - u_{k-1}) ]
                 , \label{eq:theta_update}
    \end{align}
    where 
    \begin{align}
        P_{k+1} 
            &=
                P_{k}
                -  \frac
                    { P_{k}\phi_{k-1}^\rmT  \phi_{k-1} P_{k} }
                    { 1 +   \phi_{k-1} P_{k} \phi_{k-1}^\rmT  }.
        \label{eq:P_update_noInverse}
    \end{align}
\end{proposition}

\section{Adaptive Autopilot}
\label{sec:AdaptiveAutopilot}

\begin{figure*}
    \centering
    \subfloat
    [
        PX4 autopilot interfaced with the mission planner and the quadcopter simulator. 
    ] 
    {
    \begin{tikzpicture}[auto, node distance=2cm,>=latex']
        \node [smallblock, text width=1.2cm] (Mission) { Mission Planner};
        \node [smallblock, right of=Mission,node distance=3cm, text width=1.6cm] (Pos_Cont) {Position Controller};
        \node [smallblock, right of=Pos_Cont,node distance=3.5cm,text width=1.6cm] (Att_Cont) {Attitude Controller};
        \node [smallblock, right of=Att_Cont,node distance=3cm] (Mixer) {Mixer};
        \node [smallblock, right of=Mixer,node distance=3cm] (Quadcopter) {Quadcopter};
        
        \draw [->] (Mission) -- node 
            {$\matl{c} 
                X_{\rm sp} \\ Y_{\rm sp} \\ Z_{\rm sp}
            \matr$}
            (Pos_Cont);
        \draw [->] (Pos_Cont) -- node [xshift = 0] 
            {$\matl{c} 
                F_{x,\rm sp} \\ F_{y,\rm sp} \\ F_{z\rm sp}
            \matr$}
            (Att_Cont);
        \draw [->] (Att_Cont) -- node 
            {$\matl{c} 
                \dot P_{\rm sp} \\ \dot Q_{\rm sp} \\ \dot R_{\rm sp}
            \matr$}
            (Mixer);
        \draw [->] (Mixer) -- node {$\{\omega_i\}_{i=1}^4$} (Quadcopter);
        
        \draw [->] (Quadcopter.20) -- 
                    +(1.5,0) 
                    node [xshift=0, yshift = 20] 
                    {$\matl{c} 
                        X \\ Y \\ Z
                    \matr,
                    \matl{c} 
                        U \\ V \\ W
                    \matr
                    $}
                    -- 
                    +(1.5,-3)  -| 
                    (Pos_Cont.270);
        \draw [->] (Quadcopter.340) -- 
                    +(.7,0) 
                    node [xshift=-35, yshift=-35] 
                    {$\matl{c} 
                        \Psi \\ \Theta \\ \Phi
                    \matr,
                    \matl{c} 
                        P \\ Q \\ R
                    \matr
                    $}
                    -- 
                    +(.7,-2) -| 
                    (Att_Cont.270);
        
        \draw [->] (Mission.90) -- +(0,1) node [xshift=10, yshift=10] {$\Psi_{\rm sp}$} -| (Att_Cont.90);

    \end{tikzpicture}
}
    \\
    \subfloat[Position controller]{   
    \centering
    \begin{tikzpicture}[auto, node distance=2cm,>=latex']
        \node (Ref_traj) 
            {$\matl{c} 
                X_{\rm sp} \\ Y_{\rm sp} \\ Z_{\rm sp}
            \matr$};
        \node [sum, right of=Mission,node distance=2cm] (sum1) {};
        \node [smallblock, right of=sum1,node distance=2cm] (Cont1) {$\rmP_r$};
        \node [sum, right of=Cont1,node distance=2.5cm] (sum2) {};
        \node [smallblock, right of=sum2,node distance=2cm] (Cont2) {${\rm PID}_v$};

        \draw [->] (Ref_traj) -- (sum1);
        \draw [->] (sum1) -- node [xshift=-15, yshift = -15]{$-$} (Cont1);
        \draw [->] (Cont1) -- node 
                    {$\matl{c} 
                        U_{\rm sp} \\ V_{\rm sp} \\ W_{\rm sp}
                        \matr
                    $}
                    (sum2);
        \draw [->] (sum2) -- 
                    node [xshift=-15, yshift = -15]{$-$}
                    (Cont2);
        \draw [->] (Cont2) -- 
                    +(2,0) node [xshift=20]
                    {$\matl{c} 
                        F_{x,\rm sp} \\ F_{y,\rm sp} \\ F_{z\rm sp}
                    \matr$};
        
        \draw [->] (sum1)+(0,-1.75) -- node 
            {$\matl{c} 
                        X \\ Y \\ Z
                    \matr
            $}
            (sum1.270);
        \draw [->] (sum2)+(0,-1.75) -- node 
            {$\matl{c} 
                        U \\ V \\ W
                \matr
            $}
            (sum2.270);
        
    \end{tikzpicture}
}
    \\
    \subfloat[Attitude controller]
    {
    \centering
    \begin{tikzpicture}[auto, node distance=2cm,>=latex']
        \node (Forces) 
                {$\matl{c} 
                        F_{x,\rm sp} \\ F_{y,\rm sp} \\ F_{z\rm sp}
                    \matr$};
        
        \node [smallblock, right of=Forces,node distance=2cm,text width=.75cm] (Inverter) {Static Map};
        
        \node [sum, right of=Inverter,node distance=2.5cm] (sum1) {};
        \node [smallblock, right of=sum1,node distance=1.5cm] (Cont1) {$\rmP_q$};
        
        \node [smallblock, right of=Cont1,node distance=3cm,text width=.75cm] (EAdots_to_Omega) {Static Map};

        \node [sum, right of=EAdots_to_Omega,node distance=2.75cm] (sum2) {};
        \node [smallblock, right of=sum2,node distance=1.5cm] (Cont2) {${\rm PID}_\omega$};
        \node [smallblock, above of=Cont2,node distance=2cm] (Cont_FF) {${\rm FF}_\omega$};
        \node [sum, right of=Cont2,node distance=1.5cm] (sum3) {};

        \draw [->] (Inverter)+(0,2) node [yshift=5] {$\Psi_{\rm sp}$} -- (Inverter.90) ;
        
        \draw [->] (Forces) -- (Inverter) ;
        \draw [->] (Inverter) --
                node [xshift=0] {$\matl{c} \Phi_{\rm sp} \\ \Theta_{\rm sp} \\ \Psi_{\rm sp} \matr$} (sum1);
        \draw [->] (sum1) node [xshift=7, yshift = -10]{$-$} -- (Cont1);
        \draw [->] (Cont1) -- 
                node [xshift=0] {$\matl{c} \dot \Phi_{\rm sp} \\ \dot \Theta_{\rm sp} \\ \dot \Psi_{\rm sp} \matr$} (EAdots_to_Omega);
        \draw [->] (EAdots_to_Omega) -- 
                node [xshift=-7] {$\matl{c} P_{\rm sp} \\ Q_{\rm sp} \\ R_{\rm sp} \matr$}
                (sum2);
        \draw [->] (sum2) node [xshift=7, yshift = -10]{$-$} --  (Cont2);
        \draw [->] (Cont2) -- (sum3.180);
        \draw [->] (EAdots_to_Omega.0)+(1.5,0) |- (Cont_FF.180);
        \draw [->] (Cont_FF.0) -| (sum3.90);
        
        \draw [->] (sum3.0) --  +(.5,0) node [xshift=12] 
            {$\matl{c}
                \dot P_{\rm sp} \\
                \dot Q_{\rm sp} \\
                \dot R_{\rm sp}
            \matr$}
            ;
        
        \draw [->] (sum1)+(0,-2) -- 
                node [xshift=0] {$\matl{c} \Phi \\ \Theta \\ \Psi \matr$} (sum1.270);
        \draw [->] (sum2)+(0,-2) -- 
                node [xshift=0] {$\matl{c} P \\ Q \\ R \matr$}
                (sum2.270);
        
    \end{tikzpicture}
    }
    \caption
    {
    Quadcopter simulation block diagram. 
    (a) shows the interface between the mission planner, the flight controller PX4, and the quadcopter simulator, 
    (b) shows the expanded view of the position controller, and
    (c) shows the expanded view of the attitude controller.
    }
    \label{fig:PX4_arch}
\end{figure*}
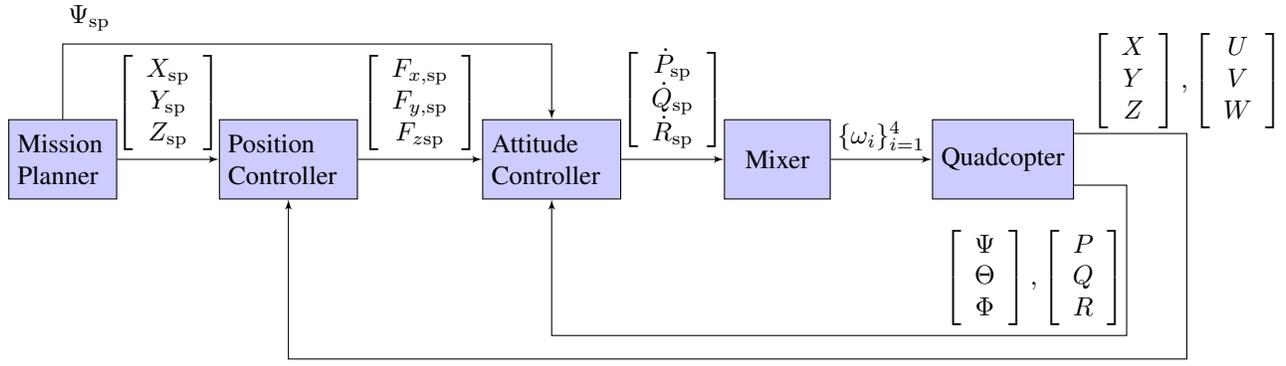
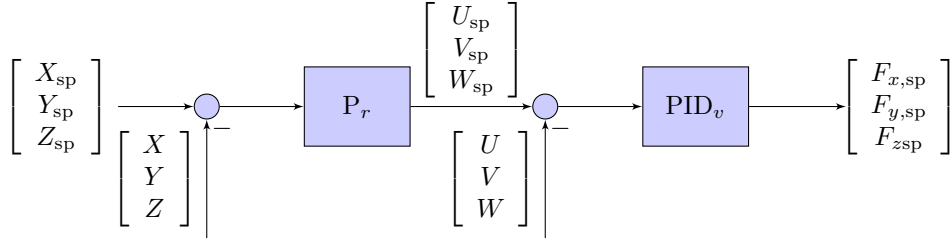
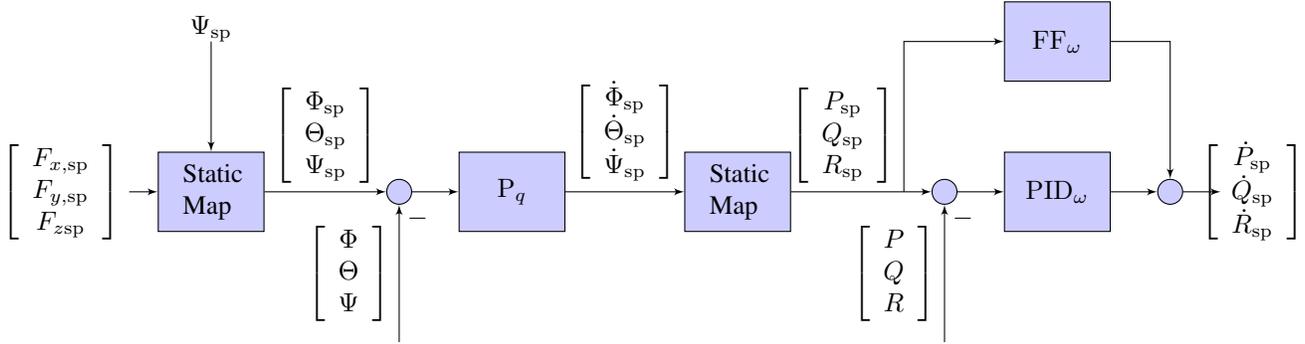

In this section, the control architecture for  flight control of a quadcopter is presented. 
The control architecture, shown in Figure \ref{fig:PX4_arch}(a), consists of a mission planner, which generates the specified trajectory, and the PX4 autopilot.
The PX4 autopilot contains a position controller in the outer loop, which generates the specified force using  position and velocity measurements, and an attitude controller in the inner loop, which generates the angular acceleration required to follow the specified trajectory using Euler-angle and angular-velocity measurements. 
Finally, depending on the geometry of the quadcopter, the angular acceleration is converted to the angular speeds $\{\omega_i\}_{i=1}^4$ of the  quadcopter motors.

The position controller shown in Figure \ref{fig:PX4_arch}(b) consists of three P controllers, which generate the specified velocities to be followed in the Earth frame using the position feedback, 
and three PID controllers, which generate the specified forces to be applied in the Earth frame using the velocity feedback.

Next, the attitude controller shown in Figure \ref{fig:PX4_arch}(c)  converts the specified forces and the specified yaw to specified Euler angles. 
The P controller generates the specified Euler angle rates, which are converted to the specified angular velocity using the appropriate orientation matrix. 
Finally, three feedforward and three PID controllers generate the angular-acceleration commands, which are converted to
angular speeds of the quadcopter motors using a static map based on the geometry of the quadcopter.


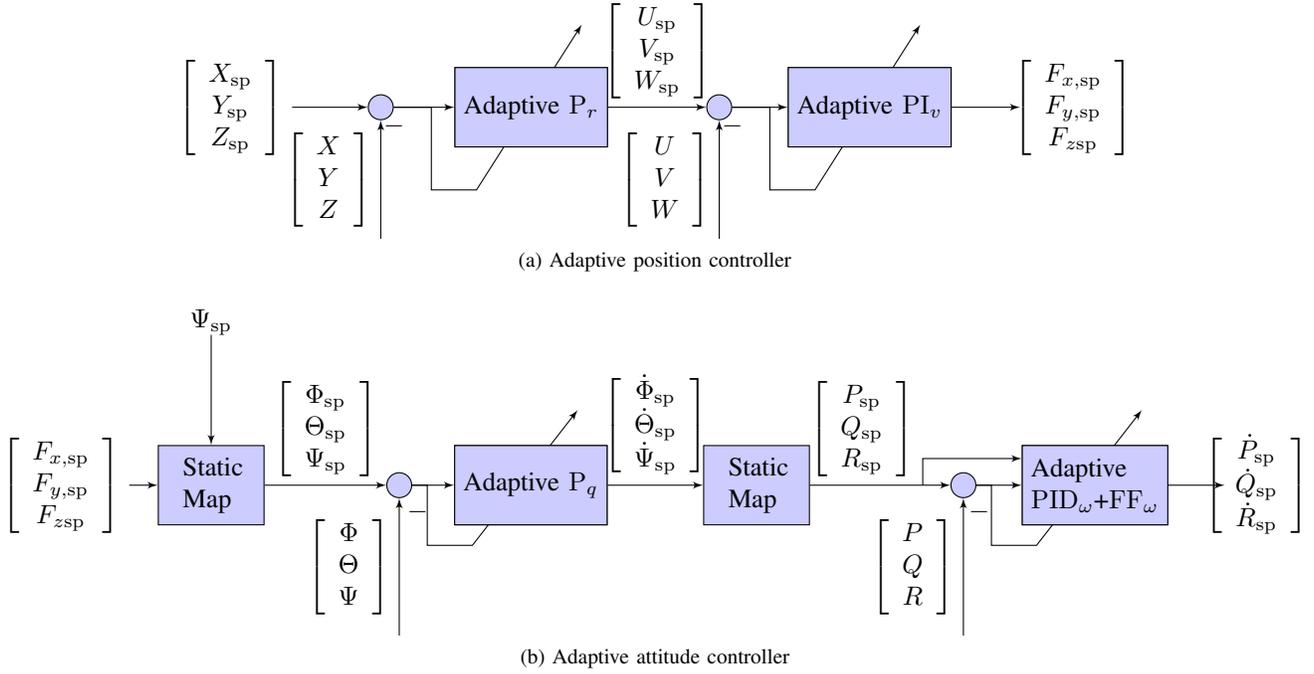
\begin{figure*}
    \centering
    \subfloat[Adaptive position controller]{   
    \centering
    \begin{tikzpicture}[auto, node distance=2cm,>=latex']
        \node (Ref_traj) 
            {$\matl{c} 
                X_{\rm sp} \\ Y_{\rm sp} \\ Z_{\rm sp}
            \matr$};
        \node [sum, right of=Mission,node distance=2cm] (sum1) {};
        \node [smallblock, right of=sum1,node distance=2cm] (Cont1) {Adaptive $\rmP_r$};
        \node [sum, right of=Cont1,node distance=2.5cm] (sum2) {};
        \node [smallblock, right of=sum2,node distance=2cm] (Cont2) {Adaptive ${\rm PI}_v$};

        \draw [->] (Ref_traj) -- (sum1);
        \draw [->] (sum1) node [xshift=5, yshift = -7]{$-$} 
                    --
                    (Cont1);
        \draw [->] (Cont1) -- node 
                    {$\matl{c} 
                        U_{\rm sp} \\ V_{\rm sp} \\ W_{\rm sp}
                        \matr
                    $}
                    (sum2);
        \draw [->] (sum2) 
                    node [xshift=5, yshift = -7]{$-$}
                    -- 
                    (Cont2);
        \draw [->] (Cont2) -- 
                    +(2,0) node [xshift=20]
                    {$\matl{c} 
                        F_{x,\rm sp} \\ F_{y,\rm sp} \\ F_{z\rm sp}
                    \matr$};
        
        \draw [->] (sum1)+(0,-1.75) -- node 
            {$\matl{c} 
                        X \\ Y \\ Z
                    \matr
            $}
            (sum1.270);
        \draw [->] (sum2)+(0,-1.75) -- node 
            {$\matl{c} 
                        U \\ V \\ W
                \matr
            $}
            (sum2.270);
        
        \draw [->] (sum2.0) -| +(.5,-1.1) -| +(1.1,-1.1) -- +(2.5,1.1);
        \draw [->] (sum1.0) -| +(.5,-1.1) -| +(1.1,-1.1) -- +(2.5,1.1);
        \node [smallblock, right of=sum2,node distance=2cm] (Cont2) {Adaptive ${\rm PI}_v$};
        \node [smallblock, right of=sum1,node distance=2cm] (Cont1) {Adaptive $\rmP_r$};
    \end{tikzpicture}
}
    \\
    \subfloat[Adaptive attitude controller]
    {
    \centering
    \begin{tikzpicture}[auto, node distance=2cm,>=latex']
        \node (Forces) 
                {$\matl{c} 
                        F_{x,\rm sp} \\ F_{y,\rm sp} \\ F_{z\rm sp}
                    \matr$};
        
        \node [smallblock, right of=Forces,node distance=2cm,text width=.75cm] (Inverter) {Static Map};
        
        \node [sum, right of=Inverter,node distance=2.5cm] (sum1) {};
        \node [smallblock, right of=sum1,node distance=1.75cm] (Cont1) {Adaptive $\rmP_q$};
        
        \node [smallblock, right of=Cont1,node distance=3cm,text width=.75cm] (EAdots_to_Omega) {Static Map};

        \node [sum, right of=EAdots_to_Omega,node distance=2.75cm] (sum2) {};
        \node [smallblock, right of=sum2,node distance=1.75cm,text width=1.7cm] (Cont2) {Adaptive ${\rm PID}_\omega$+${\rm FF}_\omega$};

        \draw [->] (Inverter)+(0,2) node [yshift=5] {$\Psi_{\rm sp}$} -- (Inverter.90) ;
        
        \draw [->] (Forces) -- (Inverter) ;
        \draw [->] (Inverter) --
                node [xshift=0] {$\matl{c} \Phi_{\rm sp} \\ \Theta_{\rm sp} \\ \Psi_{\rm sp} \matr$} (sum1);
        \draw [->] (sum1) node [xshift=7, yshift = -10]{$-$} -- (Cont1);
        \draw [->] (Cont1) -- 
                node [xshift=0] {$\matl{c} \dot \Phi_{\rm sp} \\ \dot \Theta_{\rm sp} \\ \dot \Psi_{\rm sp} \matr$} (EAdots_to_Omega);
        \draw [->] (EAdots_to_Omega) -- 
                node [xshift=-7] {$\matl{c} P_{\rm sp} \\ Q_{\rm sp} \\ R_{\rm sp} \matr$}
                (sum2);
        \draw [->] (sum2) node [xshift=6, yshift = -10]{$-$} --  (Cont2);
        \draw [->] (EAdots_to_Omega.0)+(1.5,0) |- (Cont2.160);
        
        \draw [->] (Cont2.0) --  +(.75,0) node [xshift=12] 
            {$\matl{c}
                \dot P_{\rm sp} \\
                \dot Q_{\rm sp} \\
                \dot R_{\rm sp}
            \matr$}
            ;
        
        \draw [->] (sum1)+(0,-2) -- 
                node [xshift=0] {$\matl{c} \Phi \\ \Theta \\ \Psi \matr$} (sum1.270);
        \draw [->] (sum2)+(0,-2) -- 
                node [xshift=0] {$\matl{c} P \\ Q \\ R \matr$}
                (sum2.270);
        
        \draw [->] (sum2.0) -| +(.2,-.8) -| +(.8,-.8) -- +(2.2,1.0);
        \draw [->] (sum1.0) -| +(.2,-.8) -| +(.8,-.8) -- +(2.2,1.0);
        
        \node [smallblock, right of=sum2,node distance=1.75cm,text width=1.7cm] (Cont2) {Adaptive ${\rm PID}_\omega$+${\rm FF}_\omega$};
        \node [smallblock, right of=sum1,node distance=1.75cm] (Cont1) {Adaptive $\rmP_q$};
        
    \end{tikzpicture}
    }
    \caption
    {
    Adptive PID control of Quadcopter.  
    The adaptive autopilot consists of an adaptive PID controller in the outer loop, shown in (a), and an adaptive feedforward and an adaptive PID controller in the inner loop, shown in (b).
    }
    \label{fig:PX4_arch_Adaptive}
\end{figure*}

The adaptive autopilot is constructed by modifying the PX4 autopilot.
As shown in Figure \ref{fig:PX4_arch}, the PX4 autopilot consists of three P controllers and three PID controllers in the position controller, and
three P constroller, three static-feedforward controllers, and three PID controllers in the attitude controller.

The adaptive autopilot consists of a total of twelve adaptive digital controllers, of which, six are adaptive P, three are adaptive PI, and three are adaptive PID with feedforward.
In particular, the fixed-gain P controllers $\rmP_r$ and $\rmP_q$ in the position controller and the attitude controller are replaced by the adaptive P controllers, 
the fixed-gain PID controllers ${\rm PID}_v$ in the position controller are replaced by the adaptive PI controllers, and
the fixed-gain feedforward controllers ${\rm FF}_\omega$ and the fixed-gain PID controllers ${\rm PID}_\omega$ in the attitude controller are replaced by adaptive feedforward and PID controllers. 
The modified adaptive PX4 autopilot is shown in Figure \ref{fig:PX4_arch_Adaptive}.

In the position controller, the adaptive $\rmP_r$ controller is implemented as follows. 
The error variable $z_k$ and the control $u_k$ are defined by
\begin{align}
    z_k
        \isdef 
            \matl{c}
                X_{{\rm sp}}(k \Delta t) - X(k \Delta t) \\
                Y_{{\rm sp}}(k \Delta t) - Y(k \Delta t) \\
                Z_{{\rm sp}}(k \Delta t) - Z(k \Delta t)
            \matr, 
    \
    u_k
        \isdef 
            \matl{c}
                U_{{\rm sp}}(k \Delta t) \\
                V_{{\rm sp}}(k \Delta t) \\
                W_{{\rm sp}}(k \Delta t)
            \matr, 
\end{align}
where 
$X_{{\rm sp}}(k \Delta t)$, 
$Y_{{\rm sp}}(k \Delta t)$, and
$Z_{{\rm sp}}(k \Delta t)$ are the specified positions given by the mission planner, 
$X(k \Delta t)$, 
$Y(k \Delta t)$, and
$Z(k \Delta t)$ are the measured positions, and
$U_{{\rm sp}}(k \Delta t)$,
$V_{{\rm sp}}(k \Delta t)$, and 
$W_{{\rm sp}}(k \Delta t)$ are the specified translational velocities in the inertial frame $\rm F_E$.
Note that $\Delta t = 0.04$ sec in the adaptive $\rmP_r$ controller.
Finally, for $i =1,2,3$, the control $u_{i,k}$ is given by
\begin{align}
    u_{i,k}
        =
            \theta_{i,k} z_{i,k}, 
\end{align}
where $\theta_{i,k} \in \BBR$ is given by \eqref{eq:theta_update}.
Note that the three adaptive $\rmP$ controllers in the adaptive $\rmP_r$ controller are decoupled, and hence each channel gain is independently computed using \eqref{eq:theta_update}, \eqref{eq:P_update_noInverse}.

Next, in the position controller, the adaptive ${\rm PI}_v$ controller is implemented as follows. 
The error variable $z_k$ and the control $u_k$ are defined by
\begin{align}
    z_k
        \isdef 
            \matl{c}
                U_{{\rm sp}}(k \Delta t) - U(k \Delta t) \\
                V_{{\rm sp}}(k \Delta t) - V(k \Delta t) \\
                W_{{\rm sp}}(k \Delta t) - W(k \Delta t)
            \matr, 
    \
    u_k
        \isdef 
            \matl{c}
                F_{x,{\rm sp}}(k \Delta t) \\
                F_{y,{\rm sp}}(k \Delta t) \\
                F_{z,{\rm sp}}(k \Delta t)
            \matr, 
\end{align}
where 
$U(k \Delta t)$, 
$V(k \Delta t)$, and
$W(k \Delta t)$ are the measured translational velocities in the inertial frame $\rm F_E$, and
$F_{x,{\rm sp}}(k \Delta t)$,
$F_{y,{\rm sp}}(k \Delta t)$, and 
$F_{z,{\rm sp}}(k \Delta t)$ are the specified forces in the inertial frame $\rmF_E$.
Note that $\Delta t = 0.02$ sec in the adaptive ${\rm PI}_v$ controller.
Finally, for $i =1,2,3$, the control $u_{i,k}$ is given by
\begin{align}
    u_{i,k}
        =
            \phi_{i,k} \theta_{i,k}, 
\end{align}
where $\phi_{i,k} \isdef [z_{i,k-1} \ \ \gamma_{i,k-1} ]$ and $\theta_{i,k} \in \BBR^2 $.
Note that the three adaptive $\rm PI$ controllers in the adaptive ${\rm PI}_v$ controller are decoupled, and hence the gains of each channel are independently computed using \eqref{eq:theta_update}, \eqref{eq:P_update_noInverse}.

In the attitude controller, the adaptive $\rmP_q$ controller is implemented as follows. 
The error variable $z_k$ and the control $u_k$ are defined by
\begin{align}
    z_k
        \isdef 
            \matl{c}
                \Phi_{{\rm sp}}(k \Delta t ) - \Phi(k \Delta t ) \\
                \Theta_{{\rm sp}}(k \Delta t ) - \Theta(k \Delta t ) \\
                \Psi_{{\rm sp}}(k \Delta t ) - \Psi(k \Delta t )
            \matr, 
    \
    u_k
        \isdef 
            \matl{c}
                \dot \Phi_{{\rm sp}}(k \Delta t ) \\
                \dot \Theta_{{\rm sp}}(k \Delta t ) \\
                \dot \Psi_{{\rm sp}}(k \Delta t )
            \matr, 
\end{align}
where 
$\Phi{{\rm sp}}(k \Delta t)$, 
$\Theta{{\rm sp}}(k \Delta t)$, and
$\Psi{{\rm sp}}(k \Delta t)$ are the specified Euler angles, 
$\Phi(k \Delta t)$, 
$\Theta(k \Delta t)$, and
$\Psi(k \Delta t)$ are the measured Euler angles, and
$\dot \Phi_{{\rm sp}}(k \Delta t)$,
$\dot \Theta_{{\rm sp}}(k \Delta t)$, and 
$\dot \Psi_{{\rm sp}}(k \Delta t)$ are the commanded Euler angle rates.
Note that $\Delta t = 0.004$ sec in the adaptive $\rmP_q$ controller.
Finally, for $i =1,2,3$, the control $u_{i,k}$ is given by
\begin{align}
    u_{i,k}
        =
            \theta_{i,k} z_{i,k}, 
\end{align}
where $\theta_{i,k} \in \BBR$ is given by \eqref{eq:theta_update}.
Note that the three adaptive $\rmP$ controllers in the adaptive $\rmP_q$ controller are decoupled, and hence each channel gain is independently computed using \eqref{eq:theta_update}, \eqref{eq:P_update_noInverse}.

Finally, in the attitude controller, the adaptive ${\rm PID}_\omega + {\rm FF}_\omega$ controller is implemented as follows. 
The error variable $z_k$ and the control $u_k$ are defined by
\begin{align}
    z_k
        \isdef 
            \matl{c}
                P_{{\rm sp}}(k \Delta t) - P(k \Delta t) \\
                Q_{{\rm sp}}(k \Delta t) - Q(k \Delta t) \\
                R_{{\rm sp}}(k \Delta t) - R(k \Delta t)
            \matr, 
    \
    u_k
        \isdef 
            \matl{c}
                \dot P_{{\rm sp}}(k \Delta t ) \\
                \dot Q_{{\rm sp}}(k \Delta t ) \\
                \dot R_{{\rm sp}}(k \Delta t )
            \matr, 
\end{align}
where 
$P_{{\rm sp}}(k \Delta t)$, 
$Q_{{\rm sp}}(k \Delta t)$, and
$R_{{\rm sp}}(k \Delta t)$ are the specified angular velocities in the body-fixed frame $\rm F_Q$, and
$P(k \Delta t)$, 
$Q(k \Delta t)$, and
$R(k \Delta t)$ are the measured angular velocities in the body-fixed frame $\rm F_Q$, and
$\dot P_{{\rm sp}}(k \Delta t )$,
$\dot Q_{{\rm sp}}(k \Delta t )$, and 
$\dot R_{{\rm sp}}(k \Delta t )$ are the specified angular accelerations in the inertial frame $\rmF_E$.
Note that $\Delta t = 0.004$ sec in the adaptive ${\rm PID}_\omega + {\rm FF}_\omega$ controller.
Finally, for $i =1,2,3$, the control $u_{i,k}$ is given by
\begin{align}
    u_{i,k}
        =
            \phi_{i,k} \theta_{i,k}, 
\end{align}
where 
$\phi_{i,k} \isdef [z_{i,k-1} \ \ \gamma_{i,k-1} \ \ z_{i,k-1} - z_{i,k-2} \ \ r_{i,k}]$, 
$r_k = [P_{{\rm sp}}(k \Delta t) \ Q_{{\rm sp}}(k \Delta t) \ R_{{\rm sp}}(k \Delta t)]^\rmT$, and
and $\theta_{i,k} \in \BBR^4 $.
Note that the three adaptive feedforward and PID controllers in the adaptive ${\rm PID}_\omega + {\rm FF}_\omega$ controller are decoupled, and hence the gains of each channel are independently computed using \eqref{eq:theta_update}, \eqref{eq:P_update_noInverse}.

Table \ref{tab:RCPE_variables} summarizes the variables and the hyperparameters used by RCAC in the adaptive PX4 autopilot.
\begin{table}[h]
    \centering
    \begin{tabular}{|p{2.65cm}|p{1.4cm}|c|c|c|p{.85cm}|}
        \hline
        Error variable & Control & $\theta_0$ & $\sigma$ & $P_0$ & Type  
        \\ \hline
        $X_{{\rm sp}}(k \Delta t) - X(k \Delta t)$ &
        $U_{{\rm sp}}(k \Delta t)$ &
        $0$ & $1$ & $0.01$ & P
        \\ \hline
        $Y_{{\rm sp}}(k \Delta t) - Y(k \Delta t)$ &
        $V_{{\rm sp}}(k \Delta t)$ &
        $0$ & $1$ & $0.01$ & P
        \\ \hline
        $Z_{{\rm sp}}(k \Delta t) - Z(k \Delta t)$ &
        $W_{{\rm sp}}(k \Delta t)$ &
        $0$ & $1$ & $0.01$ & P
        \\ \hline
        $\Phi_{{\rm sp}}(k \Delta t) - \Phi(k \Delta t)$ &
        $\dot \Phi_{{\rm sp}}(k \Delta t)$ &
        $0$ & $1$ & $1$ & P
        \\ \hline
        $\Theta_{{\rm sp}}(k \Delta t) - \Theta(k \Delta t)$ &
        $\dot \Theta_{{\rm sp}}(k \Delta t)$ &
        $0$ & $1$ & $1$ & P
        \\ \hline
        $\Psi_{{\rm sp}}(k \Delta t) - \Psi(k \Delta t)$ &
        $\dot \Psi_{{\rm sp}}(k \Delta t)$ &
        $0$ & $1$ & $1$ & P
        \\ \hline
        $U_{{\rm sp}}(k \Delta t) - U(k \Delta t)$ &
        $F_{x,{\rm sp}}(k \Delta t)$ &
        $0_{2 \times 1}$ & $1$ & $0.01 I_2$ & PI
        \\ \hline
        $V_{{\rm sp}}(k \Delta t) - V(k \Delta t)$ &
        $F_{y,{\rm sp}}(k \Delta t)$ &
        $0_{2 \times 1}$ & $1$ & $0.01 I_2$ & PI
        \\ \hline
        $W_{{\rm sp}}(k \Delta t) - W(k \Delta t)$ &
        $F_{z,{\rm sp}}(k \Delta t)$ &
        $0_{2 \times 1}$ & $1$ & $0.01 I_2$ & PI
        \\ \hline
        $P_{{\rm sp}}(k \Delta t) - P(k \Delta t)$ &
        $\dot P_{{\rm sp}}(k \Delta t)$ &
        $0_{4 \times 1}$ & $1$ & $0.01 I_4$ & PID+FF
        \\ \hline
        $Q_{{\rm sp}}(k \Delta t) - Q(k \Delta t)$ &
        $\dot Q_{{\rm sp}}(k \Delta t)$ &
        $0_{4 \times 1}$ & $1$ & $0.01 I_4$ & PID+FF
        \\ \hline
        $R_{{\rm sp}}(k \Delta t) - R(k \Delta t)$ &
        $\dot R_{{\rm sp}}(k \Delta t)$ &
        $0_{4 \times 1}$ & $1$ & $0.01 I_4$ & PID+FF
        \\ \hline
    \end{tabular}
    \caption{Summary of the variables and the hyperparameters used by RCAC in the adaptive PX4 autopilot.}
    \label{tab:RCPE_variables}
\end{table}

\section{Numerical Investigation}
\label{sec:Numerical}
In this section, the performance of the adaptive autopilot is investigated by numerical examples. 
In particular, the adaptive autopilot is integrated with a quadcopter simulator, where the quadcopter is commanded to reach several waypoints and return to the takeoff location. 
%
In this paper, QGroundControl is used to specify the waypoints and jMAVSim is used to simulate the quadcopter dynamics. 
3DR Iris Quadrotor airframe is selected in QGroundControl, thus setting up the controller gains and actuator constraints in PX4. 

\begin{figure}
	\centering
	\includegraphics[width=0.48\textwidth]{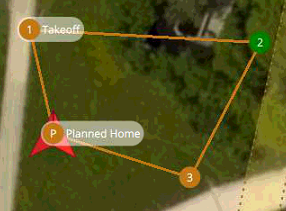}
    \caption 
    	{
    	    Top-down view of the reference trajectory specified in the mission planner.
    	}
    \label{Fig.Waypoints}
\end{figure}

\begin{figure}
	\centering
	\includegraphics[width=0.48\textwidth]{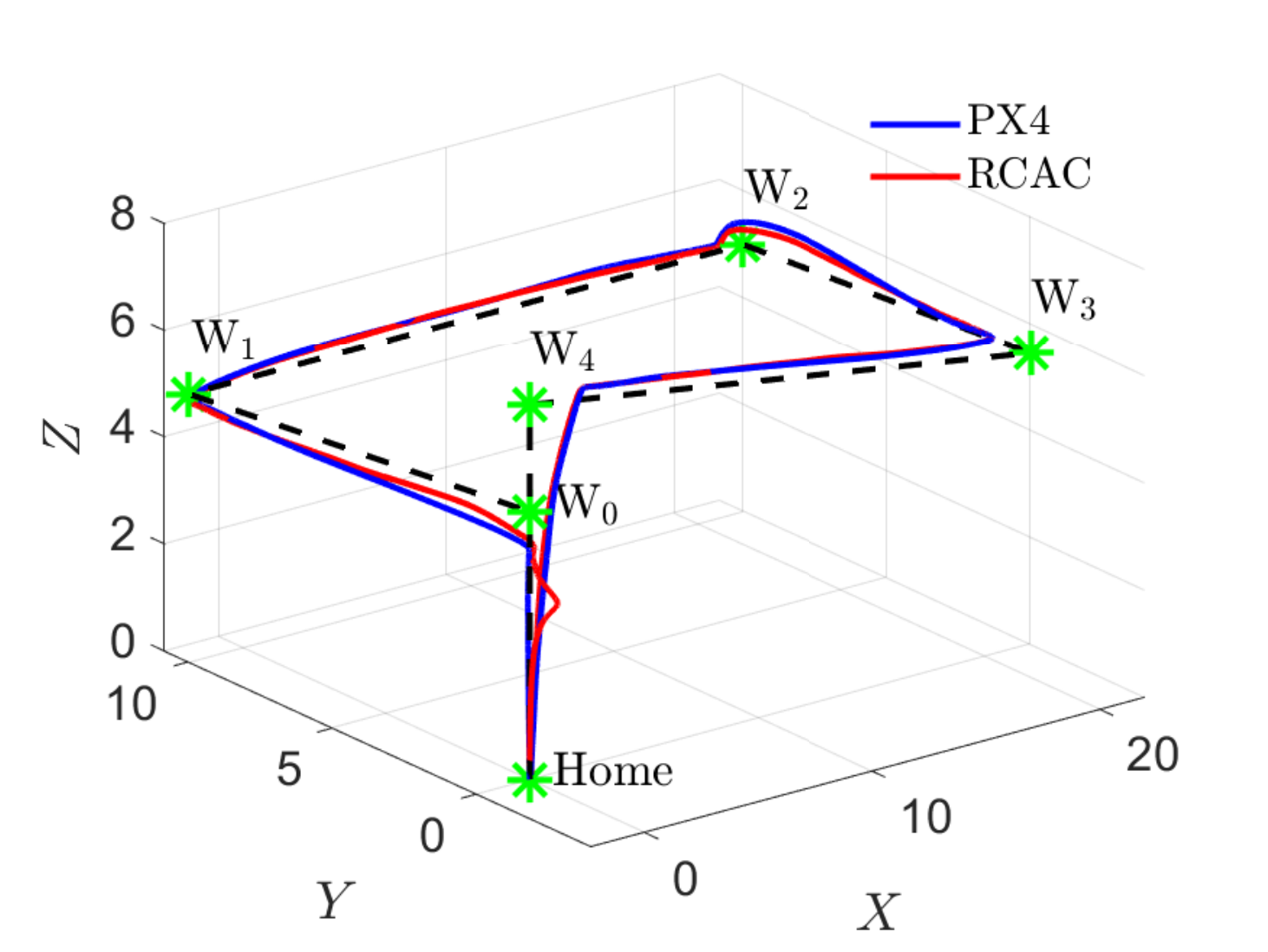}
    \caption 
    	{
    	    Closed-loop response of the quadcopter with the fixed-gain PX4 autopilot and the adaptive PX4 autopilot. 
    	    The reference trajectory is shown by the black dashes, 
    	    the quadcopter trajectory with the fixed-gain controllers is shown in blue, and
    	    the quadcopter trajectory with the adaptive controllers is shown in red.
    	    Note that all of the gains of the adaptive controllers are initialized at zero. 
    	}
    \label{Fig.Traj_Base}
\end{figure}

Figure \ref{Fig.Waypoints} shows the top-down view of the planned mission and 
Figure \ref{Fig.Traj_Base} shows the commanded trajectory in black dashes.
The command is to takeoff from Home, then fly over the waypoints $\rmW_0,$ $\rmW_1,$ $\rmW_2,$ $\rmW_3,$ and $\rmW_4,$ and finally land at Home location. 

The trajectory achieved with the default fixed-gain controllers of the PX4 autopilot is shown in blue in Figure \ref{Fig.Traj_Base}, and
the trajectory achieved with the adaptive PX4 autopilot is shown in red in Figure \ref{Fig.Traj_Base}.
Note that all adaptive controlers in the adaptive PX4 autopilot are initialized at zero. 

\begin{figure}
	\centering
	\includegraphics[width=0.48\textwidth]{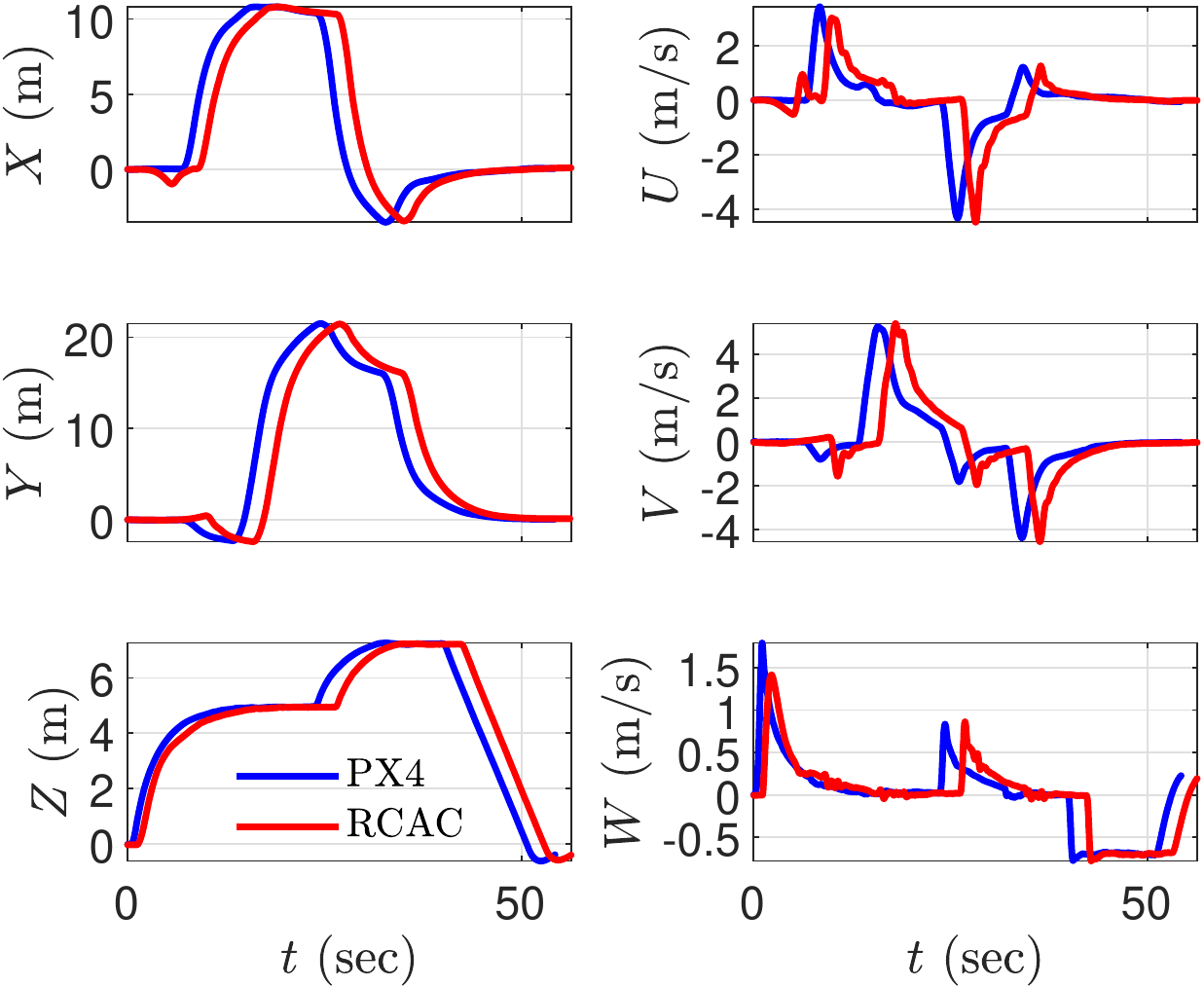}
    \caption 
    	{
    	    Closed-loop translational response of the quadcopter with the fixed-gain PX4 autopilot and the adaptive PX4 autopilot. 
    	    The quadcopter translational states with the fixed-gain controllers are shown in blue, and
    	    the quadcopter translational states with the adaptive controllers are shown in red.
    	    Note that the quadcopter response is delayed with the adaptive controllers due to the fact that all the gains of the adaptive controllers are initialized at zero.
    	}
    \label{Fig.States_Pos_Base}
\end{figure}

Figure \ref{Fig.States_Pos_Base} shows the closed-loop translational response of the quadcopter with the fixed-gain PX4 autopilot and the adaptive PX4 controller autopilot. 
The quadcopter translational states with the fixed-gain controllers are shown in blue, and
the quadcopter translational states with the adaptive controllers are shown in red.
Note that the quadcopter response is delayed with the adaptive controllers due to the fact that all the gains of the adaptive controllers are initialized at zero.

\begin{figure}
	\centering
	\includegraphics[width=0.50\textwidth, trim=15pt 0 41pt 0, clip]
	{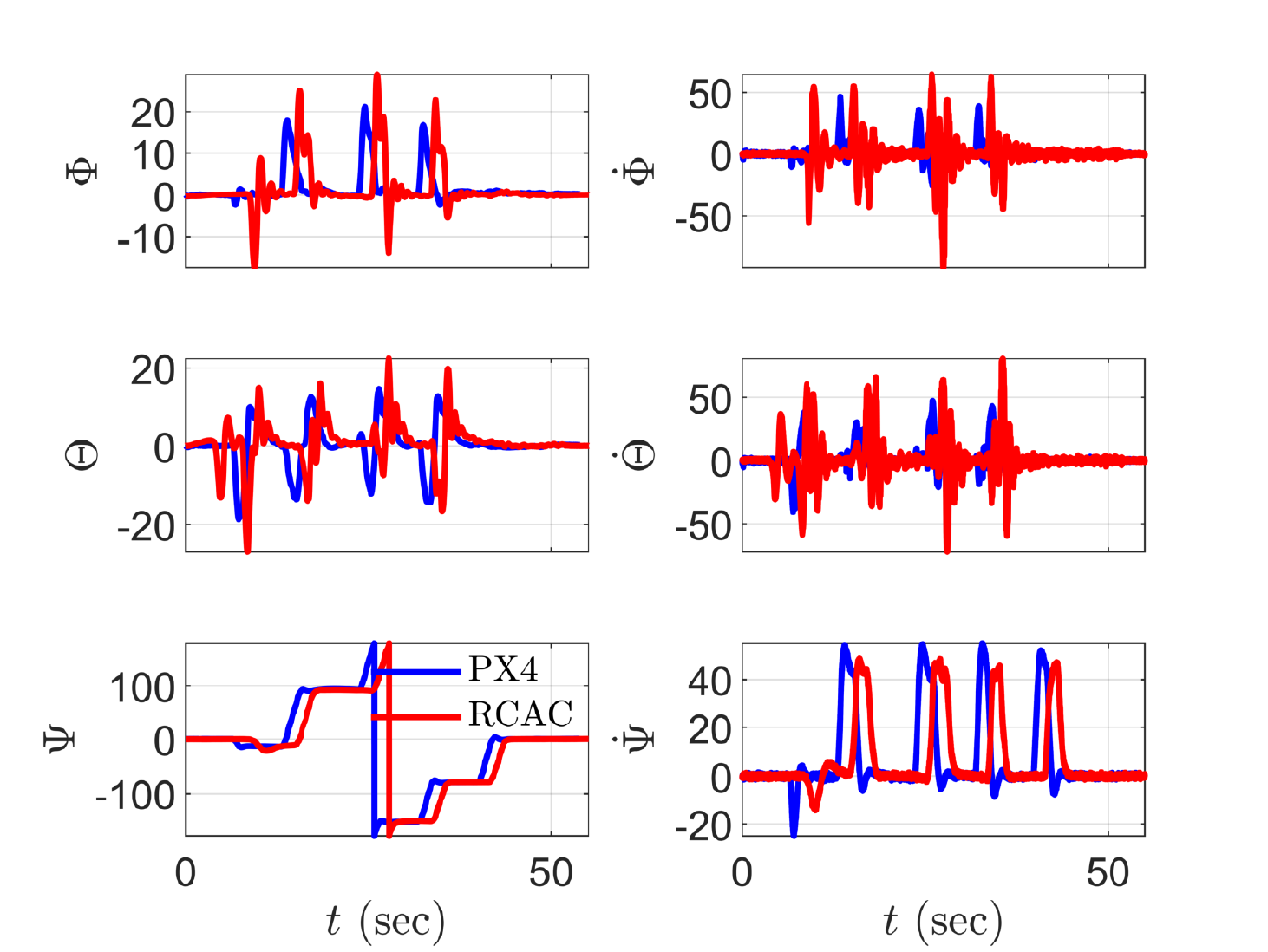}
    \caption 
    	{
    	    Closed-loop rotational response of the quadcopter with the fixed-gain PX4 autopilot and the adaptive PX4 autopilot. 
    	    The quadcopter rotational states with the fixed-gain controllers are shown in blue, and
    	    the quadcopter rotational states with the adaptive controllers are shown in red.
    	    Note that the quadcopter response is delayed with the adaptive controllers due to the fact that all of the adaptive controller gains are initialized at zero.
    	}
    \label{Fig.States_Att_Base}
\end{figure}

Figure \ref{Fig.States_Att_Base} shows the closed-loop rotational response of the quadcopter with the fixed-gain PX4 autopilot and the adaptive PX4 controller autopilot. 
The quadcopter rotational states with the fixed-gain controllers are shown in blue, and
the quadcopter rotational states with the adaptive controllers are shown in red.
Note that the quadcopter response is delayed with the adaptive controllers due to the fact that all of the adaptive controller gains are initialized at zero.

\begin{figure}
	\centering
	\includegraphics[width=0.48\textwidth]
	{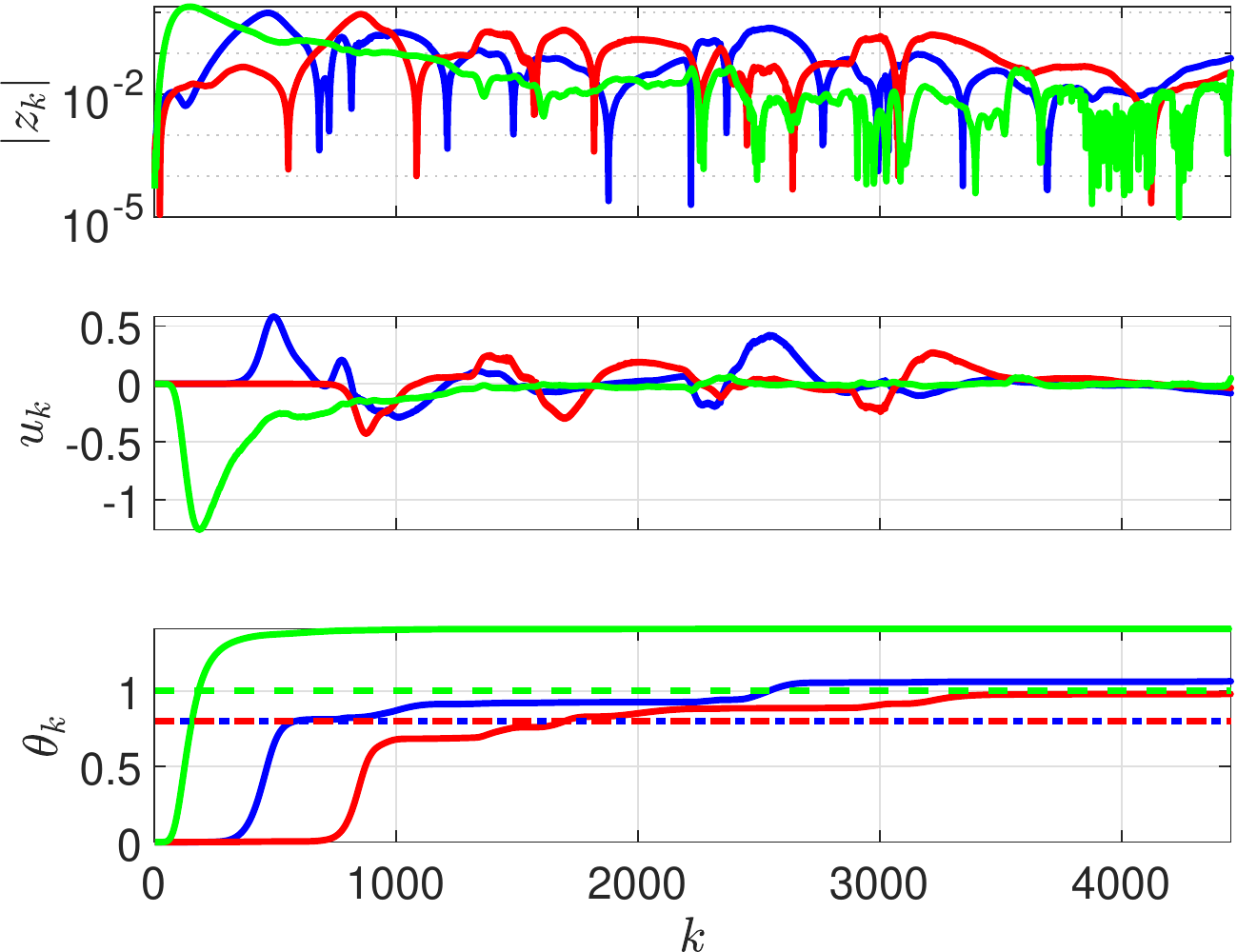}
    \caption 
    	{
    	    Adaptive $\rmP_r$ controller.
    	    The input $z_k$ to the adaptive $\rmP_r$ controller is the difference between the desired position and the measured position in the inertial frame $\rmF_\rmE$.
    	    The output $u_k$ of the adaptive $\rmP_r$ controller is the desired translational velocity in the inertial frame $\rmF_\rmE$. 
    	    The bottom-most plot shows the evolution of the adaptive proportional gains along with the default PX4 proportional gains in dashed lines.  
    	}
    \label{Fig.R_P_r}
\end{figure}

Figure \ref{Fig.R_P_r} shows the adaptive $\rmP_r$ controller variables. 
The bottom-most plot shows the evolution of the adaptive proportional gains and the corresponding stock PX4 fixed gains are shown in dashed lines. 
Note that the RCAC gains converge near the fixed gains.

\begin{figure}
	\centering
	\includegraphics[width=0.48\textwidth]
	{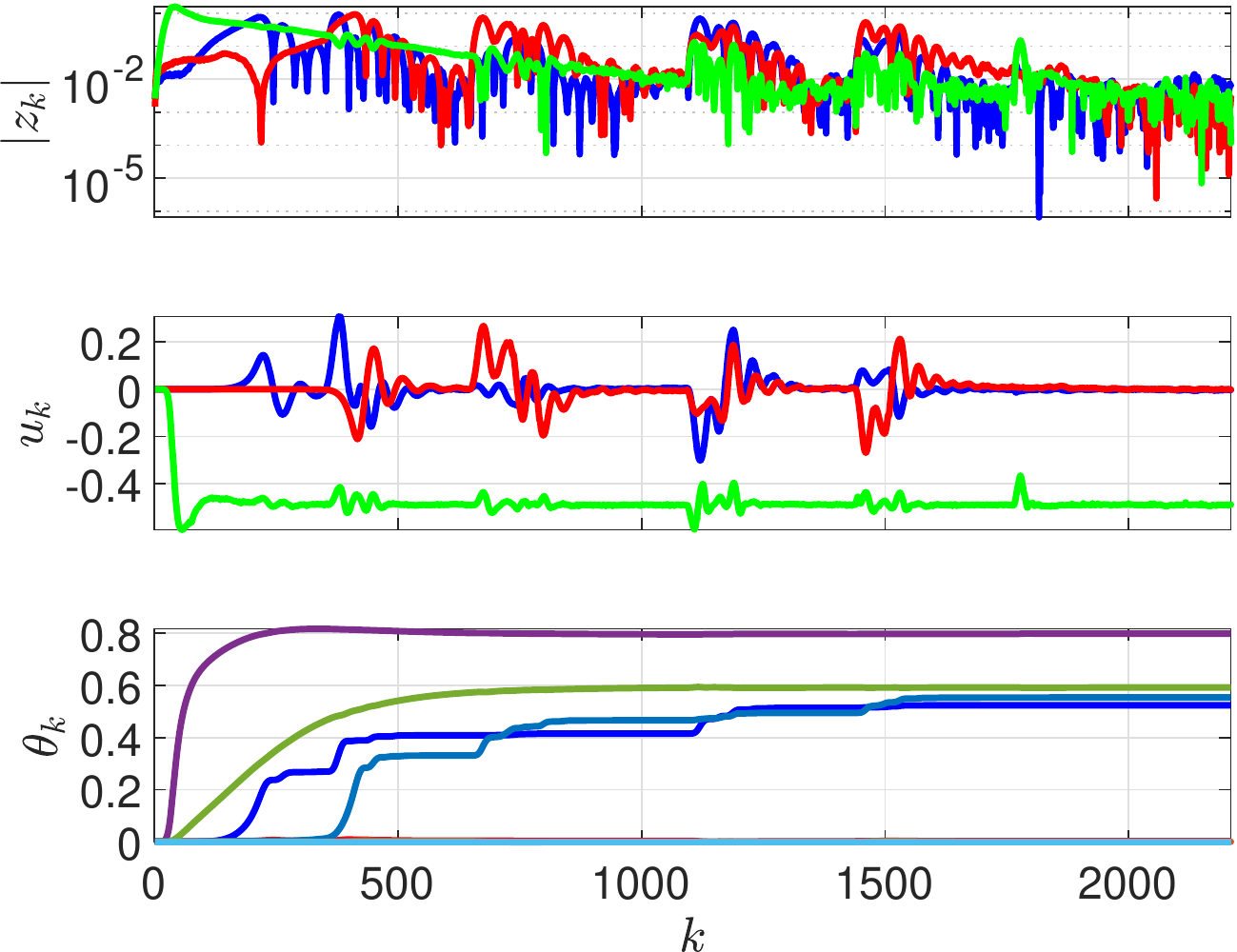}
    \caption 
    	{
    	    Adaptive ${\rm PI}_v$ controller.
    	    The input $z_k$ to the adaptive ${\rm PI}_v$ controller is the difference between the desired velocity and the measured velocity in the inertial frame $\rmF_\rmE$.
    	    The output $u_k$ of the adaptive ${\rm PI}_v$ controller is the desired force to be applied to the quadcopter along the corresponding direction. 
    	    The bottom-most plot shows the evolution of the adaptive PI gains. 
    	}
    \label{Fig.R_P_v}
\end{figure}

Figure \ref{Fig.R_P_v} shows the adaptive ${\rm PI}_v$ controller variables. 
The bottom-most plot shows the evolution of the adaptive PI gains. 

\begin{figure}
	\centering
	\includegraphics[width=0.50\textwidth, trim=13pt 0 41pt 0, clip]
	{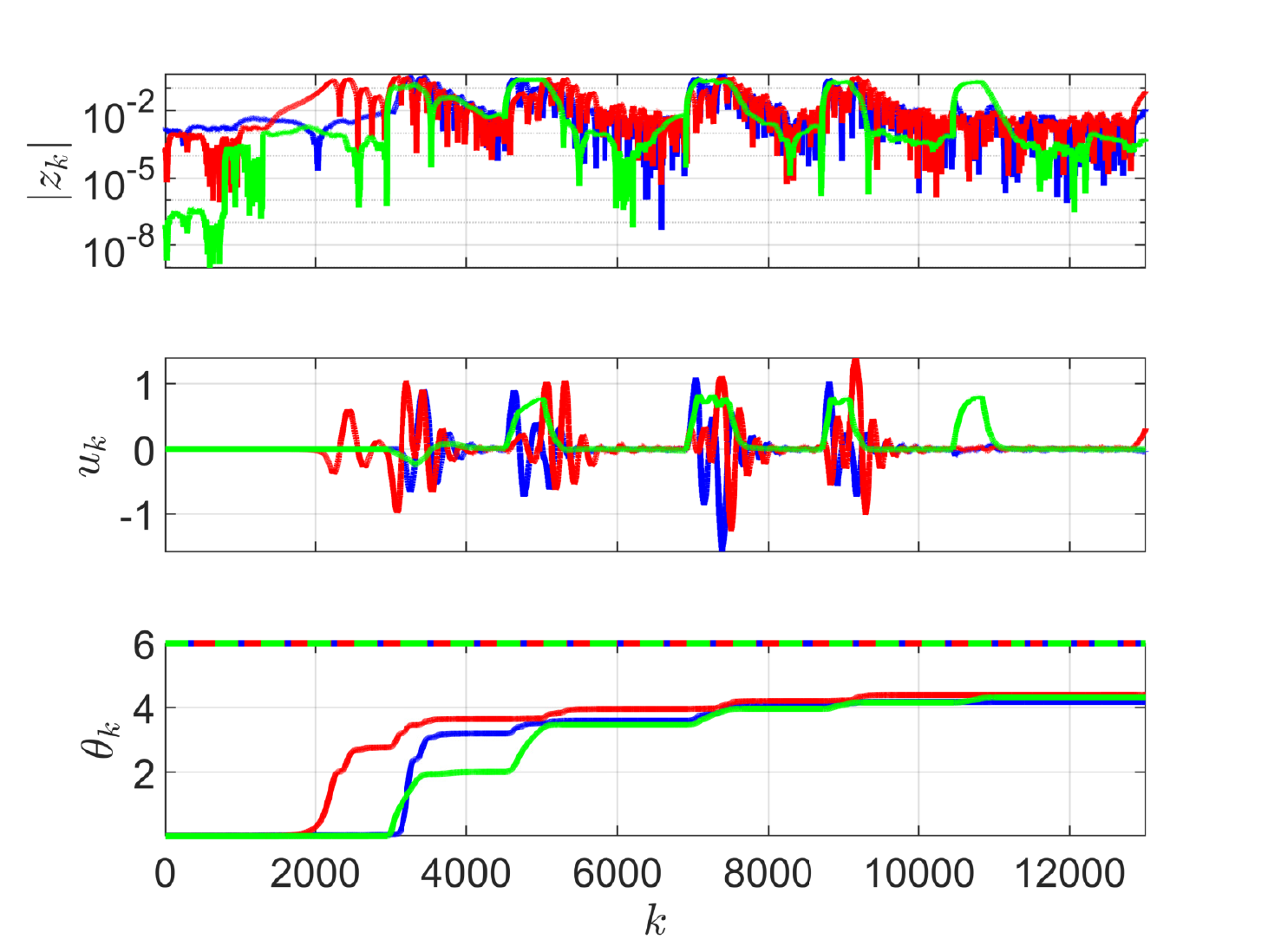}
    \caption 
    	{
    	    Adaptive ${\rm P}_q$ controller.
    	    The input $z_k$ to the adaptive ${\rm P}_q$ controller is the difference between the desired Euler angles and the measured Euler angles.
    	    The output $u_k$ of the adaptive ${\rm P}_q$ controller is the desired Euler angle rates, which are further converted into angular velocity in the body-fixed frame $\rmF_\rmQ$.  
    	    The bottom-most plot shows the evolution of the adaptive proportional gains along with the default PX4 proportional gains in dashed lines. 
    	}
    \label{Fig.R_A_q}
\end{figure}

Figure \ref{Fig.R_A_q} shows the adaptive ${\rm P}_q$ controller variables. 
The bottom-most plot shows the evolution of the adaptive P gains and the corresponding stock PX4 fixed gains are shown in dashed lines. 
Note that the RCAC gains converge near the fixed gains. 

\begin{figure}
	\centering
	\includegraphics[width=0.50\textwidth, trim=13pt 0 41pt 0, clip]
	{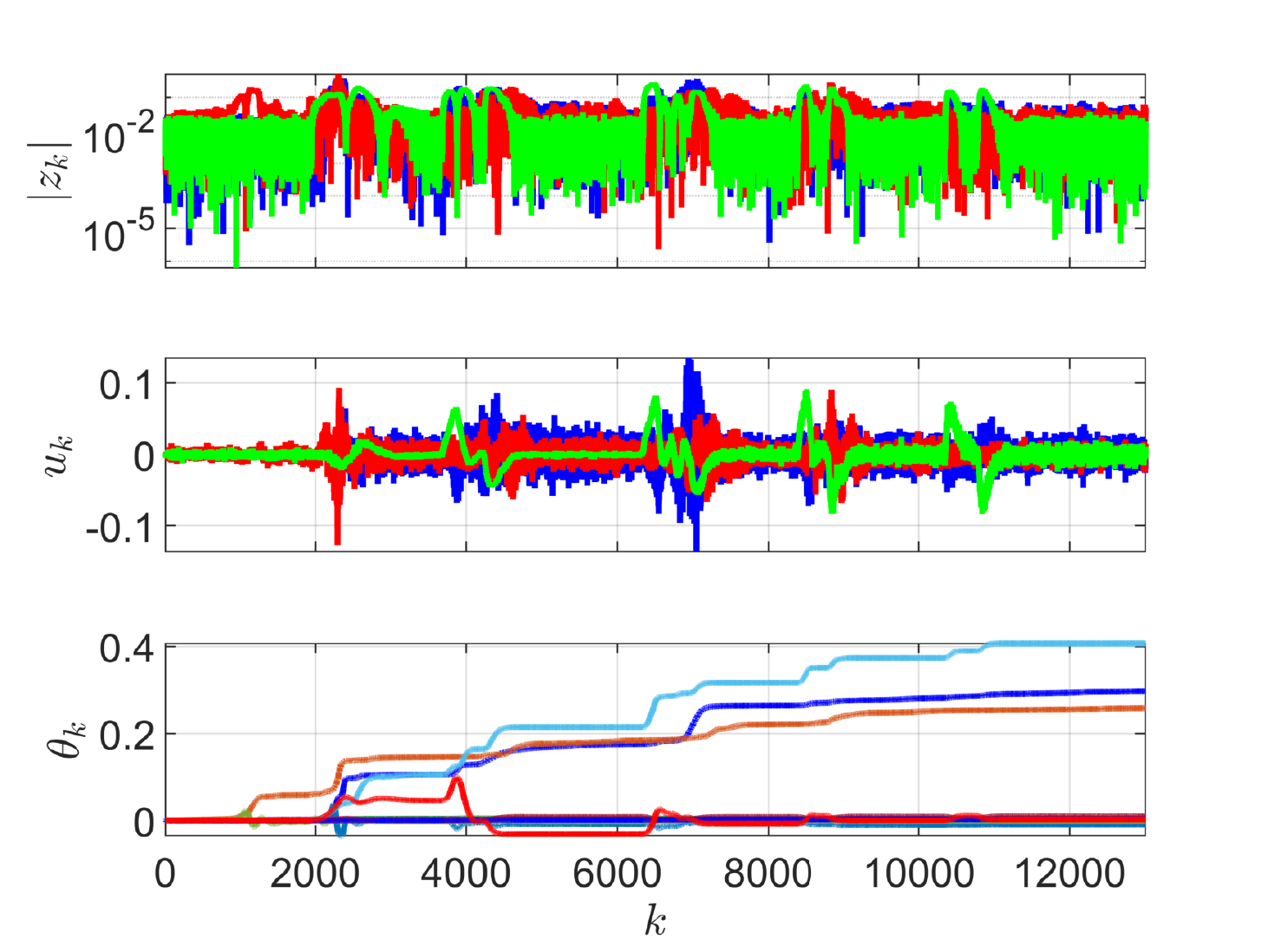}
    \caption 
    	{
    	    Adaptive ${\rm FF}_\omega+{\rm PID}_\omega$ controller.
    	    The input $z_k$ to the adaptive ${\rm FF}_\omega+{\rm PID}_\omega$ controller is the difference between the desired angular-velocity and the measured angular-velocity.
    	    The output $u_k$ of the adaptive ${\rm FF}_\omega+{\rm PID}_\omega$ controller is the desired angular-acceleration in the body-fixed frame $\rmF_\rmQ$.  
    	    The bottom-most plot shows the evolution of the adaptive feedforward and PID gains. 
    	}
    \label{Fig.R_A_w}
\end{figure}

Figure \ref{Fig.R_A_w} shows the adaptive ${\rm FF}_\omega+{\rm PID}_\omega$ controller variables. 
The bottom-most plot shows the evolution of the adaptive feedforward and the PID gains. 

\begin{figure}
	\centering
	\includegraphics[width=0.49\textwidth]
	{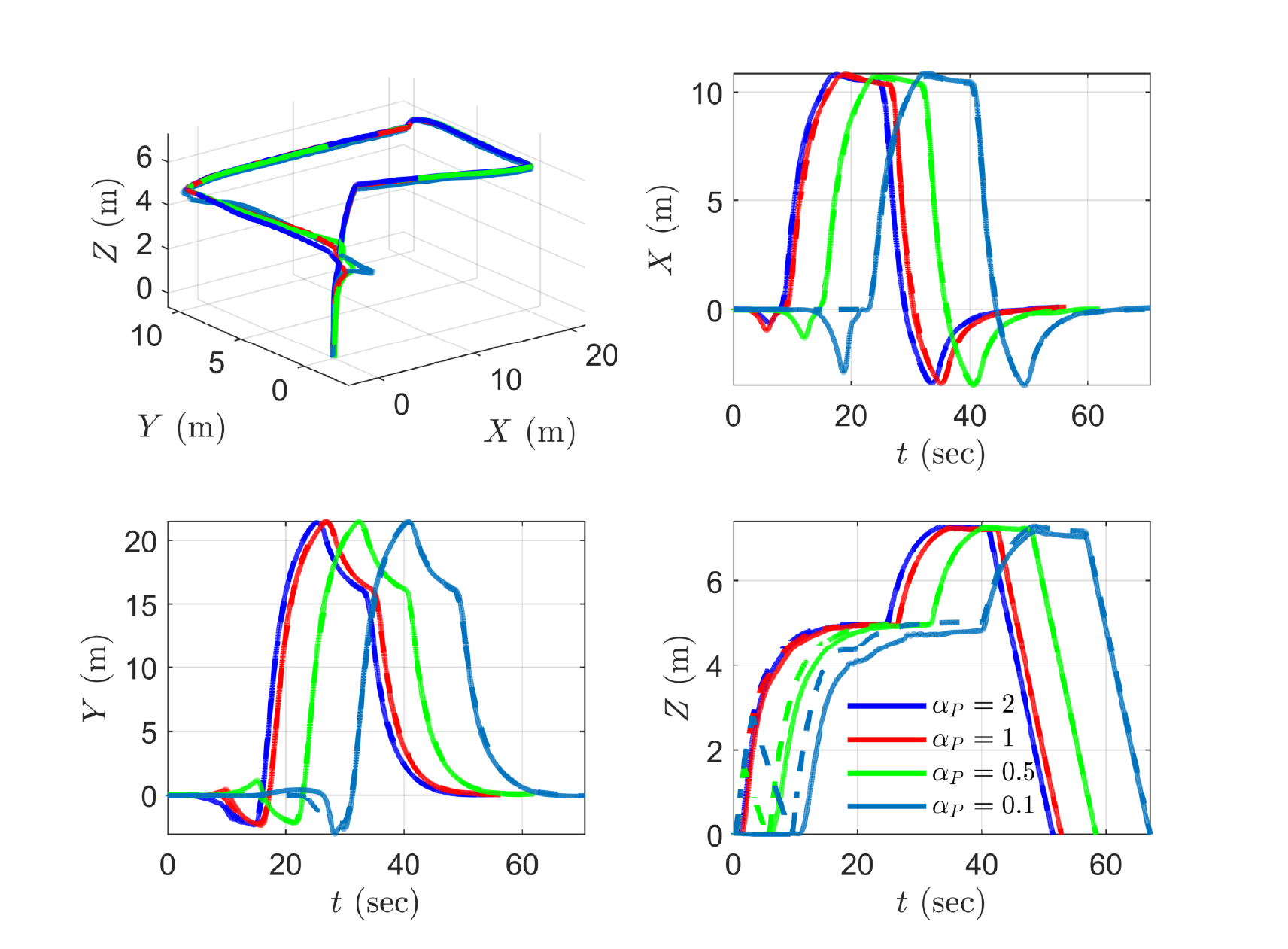}
    \caption 
    	{
    	    Effect of $P_0$ on the closed-loop response of the quadcopter with the adaptive controller.
    	}
    \label{Fig.R_alphaP}
\end{figure}

Next, the effect of the hypreparameters $P_0$ and $\sigma$ on the performance of the adaptive PX4 autopilot is investigated.
First, in all of the controllers updated by RCAC, $P_0$ is multiplied by $\alpha_P$, where $\alpha_P \in \{ 0.1, 0.5, 1, 2\}$, while all other tuning settings are held fixed. 
Figure \ref{Fig.R_alphaP} shows the trajectory achieved by the quadcopter for several values of $\alpha_P$.
Note that quadcopter response is slower for smaller values of $P_0$. 

\begin{figure}
	\centering
	\includegraphics[width=0.49\textwidth]
	{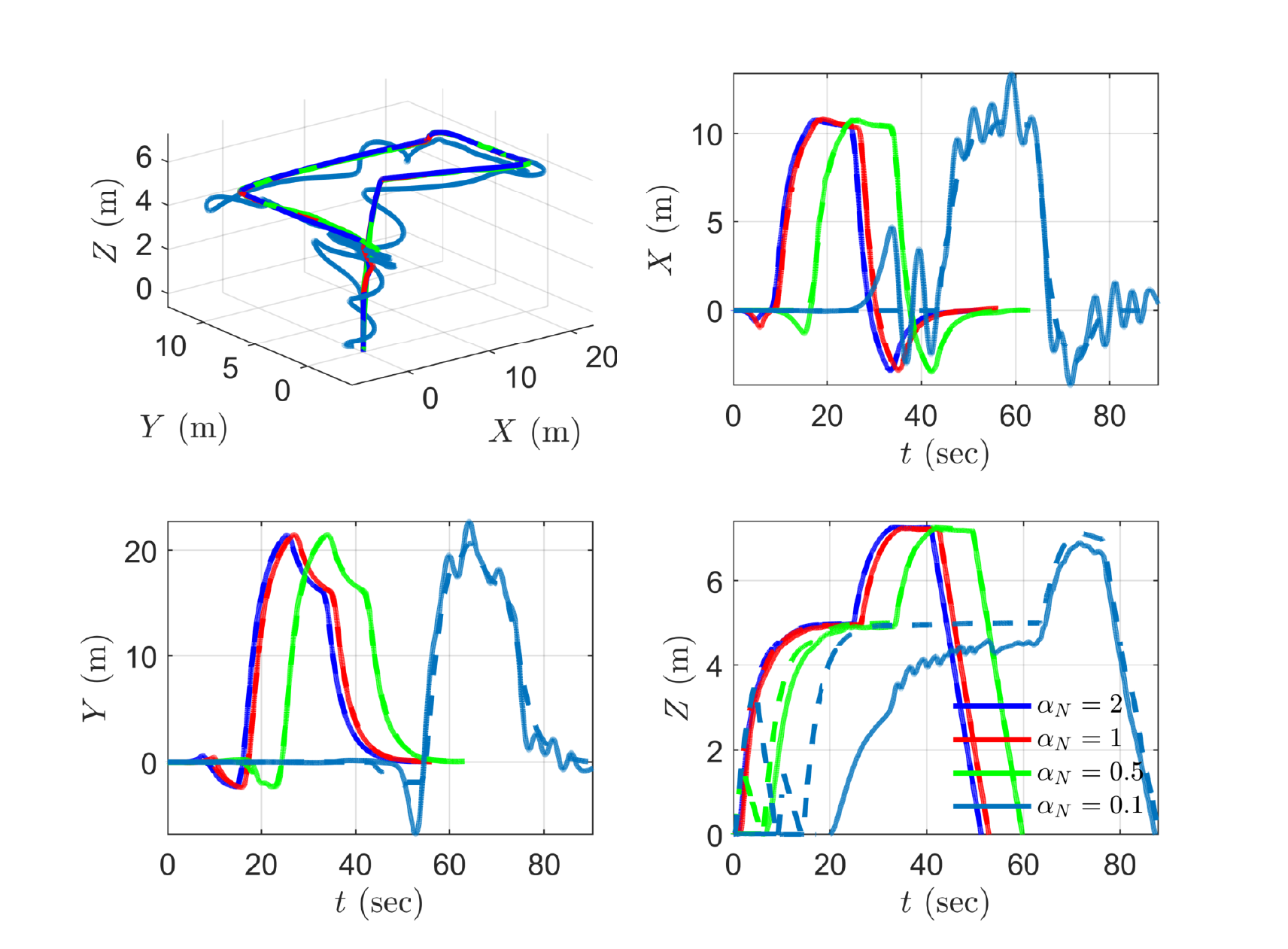}
    \caption 
    	{
    	    Effect of $\sigma$ on the closed-loop response of the quadcopter with the adaptive controller.
    	}
    \label{Fig.R_alphaN}
\end{figure}

Next, in all of the controllers updated by RCAC, $\sigma$ is multiplied by $\alpha_N$, where $\alpha_N \in \{ 0.1, 0.5, 1, 2\}$, while all other tuning settings are held fixed. 
Figure \ref{Fig.R_alphaN} shows the trajectory achieved by the quadcopter for several values of $\alpha_N$.
Note that quadcopter response degrades for smaller values of $\sigma$.

\begin{figure}
	\centering
	\includegraphics[width=0.49\textwidth]
	{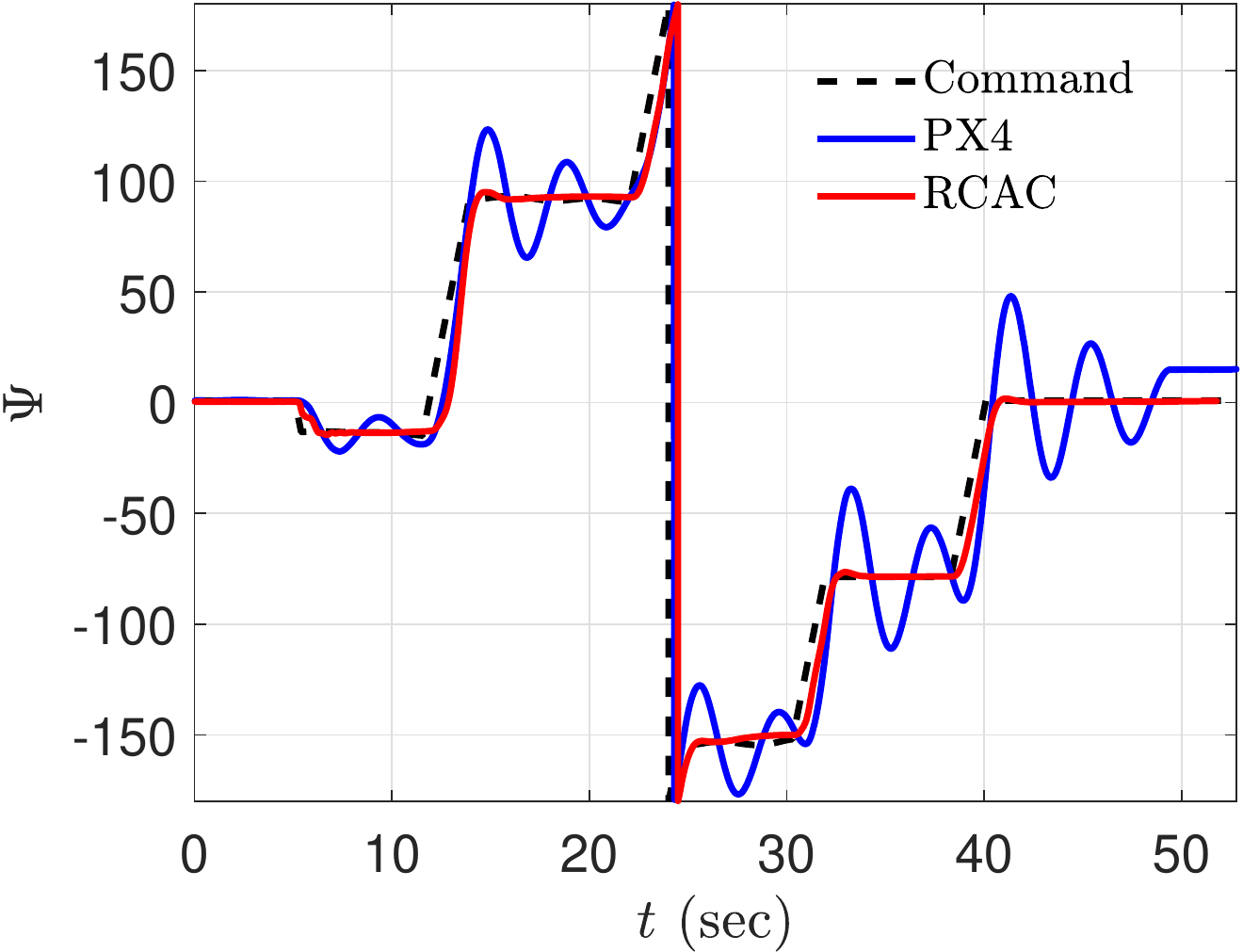}
    \caption 
    	{
    	    Closed-loop yaw response with the stock and the adaptive PX4 controller. 
    	    Note that the closed-loop yaw response with the stock PX4 controller is oscillatory, whereas the closed-loop yaw response with the adaptive PX4 controller is similar to the closed-loop yaw response obtained in the nominal condition.
    	}
    \label{Fig.Var_Quads_J5_yaw}
\end{figure}

\begin{figure}
	\centering
	\includegraphics[width=0.49\textwidth, trim=15pt 0 40pt 0, clip]
	{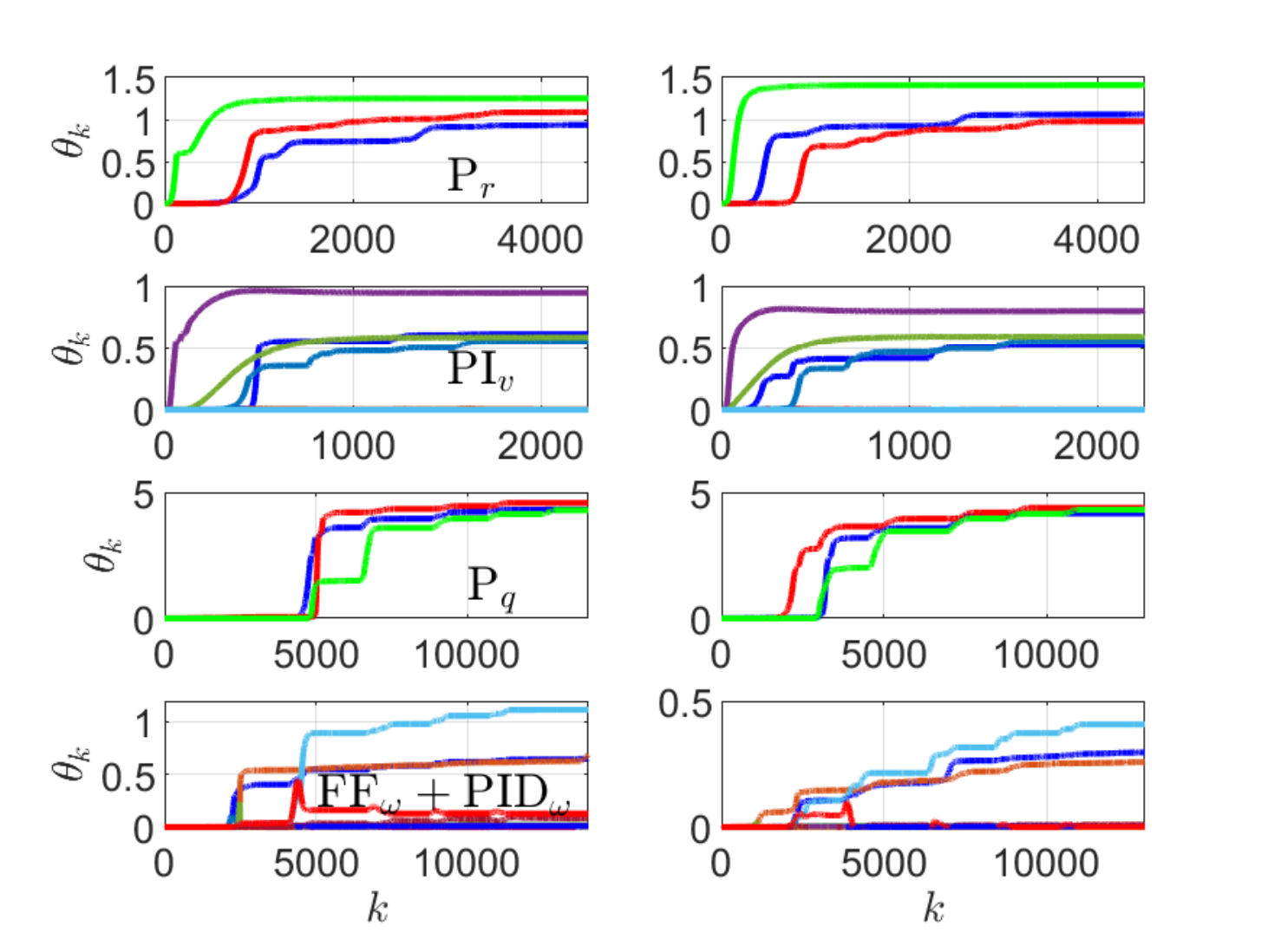}
    \caption 
    	{
    	    Adaptive PX4 controller gains in the off-nominal and the nominal condition. 
    	    Plots on the left show the controller gains in the case where the moment-of-inertia matrix of the quadcopter is scaled by a factor of five, whereas
    	    the plots on the right show the controller gains in the nominal condition.
    	    Note that the controller gains of evolve differently to maintain similar performance. 
    	}
    \label{Fig.Var_Quads_J5_1_theta}
\end{figure}

Finally, the physical properties of the simulated quadcopter are varied to compare the performance of the stock and the adaptive PX4 controller in off-nominal conditions. 
In particular, the moment-of-inertia matrix of the quadcopter is scaled by a factor of five, and  the stock and the adaptive PX4 controller are used to follow the trajectory shown in Figure \ref{Fig.Waypoints} and Figure \ref{Fig.Traj_Base}.
Figure \ref{Fig.Var_Quads_J5_yaw} shows yaw angle of the quadcopter during the trajectory. 
Note that the closed-loop yaw response with the stock PX4 controller is oscillatory, whereas the closed-loop yaw response with the adaptive PX4 controller is similar to the closed-loop yaw response obtained in the nominal condition (shown in lower-left plot in Figure \ref{Fig.States_Att_Base}).
Figure \ref{Fig.Var_Quads_J5_1_theta} shows the adaptive PX4 controller gains in the off-nominal and the nominal condition. 
Note that the controller gains of evolve accordingly to maintain similar performance. 
%
%
%

\section{Conclusions and Future Work}
\label{sec:conclusions}
This paper presented the implementation of adaptive PID controllers in the PX4 autopilot. 
In particular, the P and the PID controller in the position controller, and the P and the feedforward and PID controller in the attitude controller were replaced by adaptive controllers.
The performance of the adaptive autopilot was investigated by integrating a quadcopter simulator with the autopilot. 
The robustness of the adaptive autopilot was investigated by varying the hyperparameters of the adaptive controller and the physical moment-of-inertial matrix of the quadcopter. 
In the nominal case, the adaptive PX4 controller performance similar to the stock PX4 controller
However, in the off-nominal case, the closed-loop response of the quadcopter with the stock PX4 controller degraded considerably, whereas the adaptive PX4 controller adapted to maintain similar performance obtained in the nominal case. 

The future work will focus on investigating the performance of the retrospective-cost based adaptive PX4 controller with forgetting factor,
investigating the performance of IIR controllers in the position and attitude controllers, and
conducting physical flight tests to validate the simulation results. 


\section{Acknowledgments}
This research was supported in part by the Office of Naval Research under grant N00014-19-1-2273.  
The first author would like to thank S. A. U. Islam, G. Goel, P. S. Sharma, M. Romano, and G. Haggin for extremely helpful discussions about the PX4 autopilot. 


\bibliographystyle{IEEEtran}
\bibliography{PX4bib}

\begin{thebibliography}{10}
\providecommand{\url}[1]{#1}
\csname url@samestyle\endcsname
\providecommand{\newblock}{\relax}
\providecommand{\bibinfo}[2]{#2}
\providecommand{\BIBentrySTDinterwordspacing}{\spaceskip=0pt\relax}
\providecommand{\BIBentryALTinterwordstretchfactor}{4}
\providecommand{\BIBentryALTinterwordspacing}{\spaceskip=\fontdimen2\font plus
\BIBentryALTinterwordstretchfactor\fontdimen3\font minus
  \fontdimen4\font\relax}
\providecommand{\BIBforeignlanguage}[2]{{%
\expandafter\ifx\csname l@#1\endcsname\relax
\typeout{** WARNING: IEEEtran.bst: No hyphenation pattern has been}%
\typeout{** loaded for the language `#1'. Using the pattern for}%
\typeout{** the default language instead.}%
\else
\language=\csname l@#1\endcsname
\fi
#2}}
\providecommand{\BIBdecl}{\relax}
\BIBdecl

\bibitem{chang2016development}
C.-C. Chang, J.-L. Wang, C.-Y. Chang, M.-C. Liang, and M.-R. Lin, ``Development
  of a multicopter-carried whole air sampling apparatus and its applications in
  environmental studies,'' \emph{Chemosphere}, vol. 144, pp. 484--492, 2016.

\bibitem{anweiler2017multicopter}
S.~Anweiler and D.~Piwowarski, ``Multicopter platform prototype for
  environmental monitoring,'' \emph{Journal of Cleaner Production}, vol. 155,
  pp. 204--211, 2017.

\bibitem{andaluz2015nonlinear}
V.~H. Andaluz, E.~L{\'o}pez, D.~Manobanda, F.~Guamushig, F.~Chicaiza, J.~S.
  S{\'a}nchez, D.~Rivas, F.~P{\'e}rez, C.~S{\'a}nchez, and V.~Morales,
  ``Nonlinear controller of quadcopters for agricultural monitoring,'' in
  \emph{International Symposium on Visual Computing}.\hskip 1em plus 0.5em
  minus 0.4em\relax Springer, 2015, pp. 476--487.

\bibitem{schafer2016multicopter}
B.~E. Sch{\"a}fer, D.~Picchi, T.~Engelhardt, and D.~Abel, ``Multicopter
  unmanned aerial vehicle for automated inspection of wind turbines,'' in
  \emph{2016 24th Mediterranean Conference on Control and Automation
  (MED)}.\hskip 1em plus 0.5em minus 0.4em\relax IEEE, 2016, pp. 244--249.

\bibitem{stokkeland2015autonomous}
M.~Stokkeland, K.~Klausen, and T.~A. Johansen, ``Autonomous visual navigation
  of unmanned aerial vehicle for wind turbine inspection,'' in \emph{2015
  International Conference on Unmanned Aircraft Systems (ICUAS)}.\hskip 1em
  plus 0.5em minus 0.4em\relax IEEE, 2015, pp. 998--1007.

\bibitem{meier2015px4}
L.~Meier, D.~Honegger, and M.~Pollefeys, ``Px4: A node-based multithreaded open
  source robotics framework for deeply embedded platforms,'' in \emph{2015 IEEE
  international conference on robotics and automation (ICRA)}.\hskip 1em plus
  0.5em minus 0.4em\relax IEEE, 2015, pp. 6235--6240.

\bibitem{rahmanCSM2017}
Y.~Rahman, A.~Xie, and D.~S. Bernstein, ``{Retrospective Cost Adaptive Control:
  Pole Placement, Frequency Response, and Connections with LQG Control},''
  \emph{IEEE Contr. Sys. Mag.}, vol.~37, pp. 28--69, Oct. 2017.

\bibitem{rezaPID}
\BIBentryALTinterwordspacing
M.~Kamaldar, S.~A.~U. Islam, S.~Sanjeevini, A.~Goel, J.~B. Hoagg, and D.~S.
  Bernstein, ``Adaptive digital pid control of first-order-lag-plus-dead-time
  dynamics with sensor, actuator, and feedback nonlinearities,'' \emph{Advanced
  Control for Applications}, vol.~1, no.~1, p. e20, 2019, e20 adc2.0020.
  [Online]. Available:
  \url{https://onlinelibrary.wiley.com/doi/abs/10.1002/adc2.20}
\BIBentrySTDinterwordspacing

\bibitem{kayacan2017learning}
E.~Kayacan, M.~A. Khanesar, J.~Rubio-Hervas, and M.~Reyhanoglu, ``Learning
  control of fixed-wing unmanned aerial vehicles using fuzzy neural networks,''
  \emph{International Journal of Aerospace Engineering}, vol. 2017, 2017.

\bibitem{ansari2018retrospective}
A.~A. Ansari, N.~Zhang, and D.~Bernstein, ``Retrospective cost adaptive pid
  control of quadcopter/fixed-wing mode transition in a vtol aircraft,'' in
  \emph{2018 AIAA Guidance, Navigation, and Control Conference}, 2018, p. 1838.

\end{thebibliography}

\end{document}